\documentstyle[12pt,preprint,aps,floats]{revtex}
\def\JHEP#1#2#3{JHEP {\bf#1}, #3 (19#2)}
\def\EPJC#1#2#3{Eur. Phys. J. {\bf C#1}, #3 (19#2)}
\def\NPB#1#2#3{Nucl. Phys. {\bf B#1}, #3 (19#2)}

\def\PLB#1#2#3{Phys. Lett. {\bf B#1}, #3 (19#2)}

\def\PRD#1#2#3{Phys. Rev. {\bf D#1}, #3 (19#2)}
\def\PRL#1#2#3{Phys. Rev. Lett. {\bf#1}, #3 (19#2)}

\input epsf.tex
\def\DESepsf(#1 width #2){\epsfxsize=#2 \epsfbox{#1} \vspace*{0.07in}}
\newcommand{\postscript}[2]{\setlength{\epsfxsize}{#2\hsize}
   \centerline{\epsfbox{#1}}}

\newcommand{\tr}{\:\!\text{Tr}\:\!}
\newcommand{\sign}{\:\!\text{sign}\:\!}
\newcommand{\bsg}{b\to s \gamma}
\newcommand{\mt}{m_{\text{top}}}
\newcommand{\mgut}{M_{\text{GUT}}}
\newcommand{\maux}{M_{\text{aux}}}

\newcommand{\tb}{\tan\beta}
\newcommand{\stau}{\tilde{\tau}}
\newcommand{\selectron}{\tilde{e}}
\newcommand{\chargino}{\tilde{\chi}^{\pm}}
\newcommand{\neutralino}{\tilde{\chi}^0}
\newcommand{\hplus}{H^{\pm}}
\newcommand{\amu}{a_{\mu}^{\text{SUSY}}}
\newcommand{\bold}[1]{\mbox{\boldmath $#1$}}

\newcommand{\ifb}{ \text{ fb}^{-1}}
\newcommand{\tev}{\text{ TeV}}
\newcommand{\gev}{\text{ GeV}}
\newcommand{\mev}{\text{ MeV}}

\tighten

\begin{document}

\preprint{
\noindent
\begin{minipage}[t]{3in}
\begin{flushleft}
July 1999 \\
\end{flushleft}
\end{minipage}
\hfill
\begin{minipage}[t]{3in}
\begin{flushright}
IASSNS--HEP--99--65\\
hep-ph/9907319\\
\vspace*{.7in}
\end{flushright}
\end{minipage}
}

\title{Supernatural Supersymmetry: \\
Phenomenological Implications of\\
Anomaly-Mediated Supersymmetry Breaking}

\author{Jonathan L. Feng and Takeo Moroi
\vspace*{.2in}
}
\address{School of Natural Sciences,
Institute for Advanced Study\\ Princeton, NJ 08540 USA
\vspace*{.2in}
}

\maketitle

\begin{abstract}
We discuss the phenomenology of supersymmetric models in which
supersymmetry breaking terms are induced by the super-Weyl anomaly.
Such a scenario is envisioned to arise when supersymmetry breaking
takes place in another world, i.e., on another brane. We review the
anomaly-mediated framework and study in detail the minimal
anomaly-mediated model parametrized by only $3+1$ parameters: $\maux$,
$m_0$, $\tb$, and $\sign(\mu)$.  The renormalization group equations
exhibit a novel ``focus point'' (as opposed to fixed point) behavior,
which allows squark and slepton masses far above their usual
naturalness bounds.  We present the superparticle spectrum and
highlight several implications for high energy colliders.  Three
lightest supersymmetric particle (LSP) candidates exist: the Wino, the
stau, and the tau sneutrino. For the Wino LSP scenario, light Wino
triplets with the smallest possible mass splittings are preferred;
such Winos are within reach of Run II Tevatron searches.  Finally, we
study a variety of sensitive low energy probes, including $\bsg$, the
anomalous magnetic moment of the muon, and the electric dipole moments
of the electron and neutron.
\end{abstract}

\pacs{PACS numbers: 14.80.Ly, 11.30.Er, 12.60.Jv, 11.30.Pb}


\section{Introduction}
\label{sec:intro}

Signals from supersymmetry (SUSY) are important targets for particle
physics experiments.  These signals range from the direct discovery of
supersymmetric particles at high energy colliders to indirect signals
at lower energy experiments through measurements of flavor-changing
processes, magnetic and electric dipole moments, and so on.  The set
of possible signals and the promise of individual experiments for SUSY
searches depend strongly on what model of SUSY breaking is assumed.
It is therefore important to understand the characteristic features
and predictions of well-motivated SUSY breaking scenarios.

Probably the most well-known scenario is that of SUSY breaking in the
supergravity framework, i.e., ``gravity-mediated'' SUSY breaking.  In
this framework, SUSY breaking originates in a hidden sector and is
transmitted to the observable sector though Planck scale-suppressed
operators.  In particular, soft masses for squarks, sleptons, and
Higgs bosons are induced by direct K\"ahler interactions between
hidden and observable sector fields.  Unfortunately, these K\"ahler
interactions are not, in general, flavor-diagonal. Squark and slepton
mass matrices therefore typically have large flavor mixings, and these
induce unacceptably large flavor-changing processes, such as
$K^0$-$\bar{K}^0$ mixing and $\mu\to e\gamma$~\cite{NPB447-321}.
These difficulties, together commonly referred to as the SUSY flavor
problem, may be avoided if the K\"ahler potential is somehow
constrained to be flavor-diagonal.  Gauge-mediated SUSY
breaking~\cite{GMSB} is one proposal for solving this problem.

Recently the mechanism of ``anomaly-mediated'' SUSY breaking has been
proposed as a possibility for generating (approximately)
flavor-diagonal squark and slepton mass matrices~\cite{RS}.  In this
scenario, SUSY is again broken in a hidden sector, but it is now
transmitted to the observable sector dominantly via the super-Weyl
anomaly~\cite{RS,GLMR}.  Gaugino and scalar masses are then related to
the scale dependence of the gauge and matter kinetic functions.  For
first and second generation fields, whose Yukawa couplings are
negligible, wavefunction renormalization is almost completely
determined by gauge interactions.  Their anomaly-mediated soft scalar
masses are thus almost diagonal, and the SUSY flavor problem is
solved.  Note that this solution requires that the anomaly-mediated
terms be the dominant contributions to the SUSY breaking parameters.
This possibility may be realized, for example, if SUSY breaking takes
place in a different world, i.e., on a brane different from the
3-brane of our world, and direct K\"ahler couplings are thereby
suppressed~\cite{RS}.

As will be discussed below, the expressions for anomaly-mediated SUSY
breaking terms are scale-invariant.  Thus, they are completely
determined by the known low energy gauge and Yukawa couplings and an
overall mass scale $\maux$.  Anomaly-mediated SUSY breaking is
therefore highly predictive, with fixed mass ratios motivating
distinctive experimental signals, such as macroscopic tracks from
highly degenerate Wino-like lightest supersymmetric particles
(LSPs)~\cite{FMRSS,GGW}.  Unfortunately, one such prediction, assuming
minimal particle content, is that sleptons are tachyons.  Several
possible solutions to this problem have already been
proposed~\cite{RS,PR,Luty,KSS}. We will adopt a phenomenological
approach, first taken in Ref.~\cite{GGW}, and assume that the
anomaly-mediated scalar masses are supplemented by an additional
universal contribution $m_0^2$.  For large enough $m_0$, the slepton
squared masses are positive.  Along with the requirement of proper
electroweak symmetry breaking, this defines the minimal
anomaly-mediated model in terms of only 3+1 parameters: $\maux$,
$m_0$, $\tb$, and $\sign (\mu)$, where $\tb$ is the ratio of Higgs
vacuum expectation values (VEVs), and $\mu$ is the Higgsino mass
parameter.  The simplicity of this model allows one to thoroughly
examine all of parameter space.

In this paper, we present a detailed study of the phenomenology of the
minimal anomaly-mediated model.  We begin in Sec.~\ref{sec:AMSB} with
a brief discussion of the mechanism of anomaly-mediated SUSY breaking.
In Sec.~\ref{sec:MAM} we review the tachyonic slepton problem and the
universal $m_0$ ``solution,'' and present in detail the minimal
anomaly-mediated model described above.  The universal scalar mass
$m_0$ breaks the simple scale invariance of expressions for soft
terms.  However, this breaking is rather minimal, in a sense to be
explained, and the minimal anomaly-mediated model inherits several
simple properties from the pure anomaly-mediated case.

The naturalness of this model is examined in
Sec.~\ref{sec:naturalness}.  We find that the minimal anomaly-mediated
model exhibits a novel renormalization group (RG) ``focus point'' (as
opposed to fixed point) behavior, which allows slepton and squark
masses to be well above their usual naturalness bounds.  The title
``supernatural supersymmetry'' derives from this feature and the
envisioned other-worldly SUSY breaking.

We then turn in Sec.~\ref{sec:highenergy} to high-energy experimental
implications.  We explore the parameter space and find a variety of
interesting features, including 3 possible LSP candidates: a
degenerate triplet of Winos, the lighter stau $\stau_1$, and the tau
sneutrino $\tilde{\nu}_{\tau}$.  The Wino LSP scenario is realized in
a large fraction of parameter space and has important new implications
for both collider physics~\cite{FMRSS,GGW} and
cosmology~\cite{GGW,MR}.  We find that naturalness and electroweak
symmetry breaking favor light Winos with the smallest possible mass
splittings, i.e., the ideal region of parameter space for Wino
searches and within the discovery reach of Run II of the Tevatron.

While anomaly-mediated models have the virtue that they predict very
little flavor-changing in the first and second generations, they are
not therefore automatically safe from all low-energy probes.  In
Sec.~\ref{sec:lowenergy} we analyze several sensitive low-energy
processes: $\bsg$, which probes flavor-changing in the third
generation, and three important flavor-{\em conserving} observables,
the anomalous magnetic dipole moment of the muon, and the electric
dipole moments of the electron and neutron.

Our conclusions and final remarks are collected in
Sec.~\ref{sec:conclusions}.  In the Appendix, we present expressions
for anomaly-mediated SUSY breaking terms in a general supersymmetric
theory and also the full flavor-dependent expressions for the specific
case of the minimal anomaly-mediated model.

\section{Anomaly-Mediated Supersymmetry Breaking}
\label{sec:AMSB}

In supergravity, SUSY breaking parameters always receive
anomaly-mediated contributions.  However, in the usual
gravity-mediated SUSY breaking scenario, SUSY breaking masses also
arise from direct interactions of observable sector fields with hidden
sector SUSY breaking fields.  Such contributions are usually
comparable to the gravitino mass, and so anomaly-mediated
contributions, which are loop-suppressed relative to the gravitino
mass, are sub-leading.  However, in a model with no direct coupling
between observable and hidden sectors, the anomaly-mediated terms can
be the dominant contributions.  In this paper, we assume that this is
the case, and that the anomaly-mediated terms are (one of) the leading
contributions to the SUSY breaking parameters.  This is realized, for
example, in the ``sequestered sector'' model of Ref.~\cite{RS}, where
the SUSY breaking sector and the observable sector are assumed to lie
on different branes, thereby suppressing direct observable
sector-hidden sector couplings.

In global SUSY, the (loop-corrected) effective Lagrangian may be
written as

\begin{eqnarray}
  {\cal L}_{\rm global} (\Box,\Lambda_{\rm cut}^*,\Lambda_{\rm cut}) 
  &=& \frac{1}{4} \int d^2\theta 
  \left[ \frac{1}{g^2} 
  - \frac{b}{8\pi^2}\log(\Box^{1/2}/\Lambda_{\rm cut}) \right] 
  W^\alpha W_\alpha + {\rm h.c.}
 \nonumber \\ &&
  + \int d^4\theta Z_\phi (\Box, \Lambda_{\rm cut}^*\Lambda_{\rm cut}) 
  \phi^* \phi
 \nonumber \\ &&
  + \int d^2\theta Y \phi^3 + {\rm h.c.}
  + \cdots \ ,
\label{Lglobal}
\end{eqnarray}
where $W^\alpha$ and $\phi$ are the gauge field strength and chiral
superfields, respectively.  Here $b$ is the $\beta$-function
coefficient for the gauge coupling constant $g$, $Z_\phi$ is the
wavefunction renormalization factor of $\phi$, $Y$ is the Yukawa
coupling constant, and $\Lambda_{\rm cut}$ is the cut-off of the
theory.

However, once we consider local SUSY, i.e., supergravity, this
expression is modified.  The most important modification for our
argument results from the fact that, in global SUSY, $\Box$ is given
by $g^{\mu\nu}\partial_\mu\partial_\nu$.  In supergravity,
$g^{\mu\nu}$ becomes a dynamical field and is part of the supergravity
multiplet. $\Box$ must therefore be promoted to an object compatible
with supergravity.  The complete expression for $\Box$ is complicated.
However, since we are interested only in the SUSY breaking terms, our
task is simplified. Perhaps the easiest prescription for deriving the
SUSY breaking terms is to introduce the compensator superfield $\Phi$,
whose VEV is given by

\begin{eqnarray}
\langle\Phi\rangle = 1 - \maux \theta^2 \ .
\label{<Phi>}
\end{eqnarray}
Here $M_{\rm aux}$ is proportional to the VEV of an auxiliary field in
the supergravity multiplet and is of order the gravitino mass after
SUSY breaking.  With this compensator field, all of the terms relevant
for calculating the anomaly-mediated SUSY breaking parameters are
contained in the Lagrangian~\cite{RS,GLMR}~\footnote{We assume there
are no Planck scale VEVs.}

\begin{eqnarray}
  {\cal L}_{\rm SUGRA} \simeq
  {\cal L}_{\rm global} 
  (\Box, \Lambda_{\rm cut}^*\Phi^*, \Lambda_{\rm cut}\Phi)\ .
\end{eqnarray}
Because $\Box$ appears in Eq.~(\ref{Lglobal}) only through terms
$\Box^{1/2}/\Lambda_{\rm cut}$ and $\Box^{1/2}/\Lambda_{\rm cut}^*$,
the replacement of $\Box$ by its supergravity generalization is
effectively carried out by the replacement $\Lambda_{\rm cut} \to
\Lambda_{\rm cut}\Phi$~\cite{PR}.

Expanding the above Lagrangian with the VEV of $\Phi$ given in
Eq.~(\ref{<Phi>}), and solving the equation of motion for the
auxiliary component of $\phi$, the anomaly-mediated contributions to
the gaugino mass $M_\lambda$, scalar squared mass $m^2$, and trilinear
scalar coupling $A$ are

\begin{eqnarray}
  M_\lambda |_{\rm AM} &=& \frac{1}{16\pi^2} b g^2 \maux
 \label{M_G(AMSB)}
 \\
  m^2|_{\rm AM} &=& \frac{1}{2} 
  \dot{\gamma} \maux^2
 \\
  A|_{\rm AM} &=& - \sum Y \gamma \maux \ ,
 \end{eqnarray}
 where
\begin{equation}
\gamma \equiv - \frac{1}{2} \frac{dZ_\phi}{d\log\Box^{1/2}} \ , \quad
\dot{\gamma} \equiv \frac{d\gamma}{d\log\Box^{1/2}} \ .
\label{A(AMSB)}
\end{equation}
Here $b$ and $\gamma$ are to be evaluated with the supersymmetric
field content present at the appropriate scale. In the above formulae,
indices have been suppressed.  The full expressions for general chiral
superfield content may be found in the Appendix.

One important feature of this result is that the formulae for the
anomaly-mediated SUSY breaking parameters are
RG-invariant~\cite{RS,GLMR,PR,JJ}.  The anomaly-induced masses are
given as functions of the gauge and Yukawa coupling constants, as
shown in Eqs.~(\ref{M_G(AMSB)}) -- (\ref{A(AMSB)}), and the
$\beta$-functions for the individual SUSY breaking parameters agree
with the $\beta$-functions of the right-hand sides whose scale
dependences are determined through the gauge and Yukawa coupling RG
equations.

\section{The Minimal Anomaly-mediated Model}
\label{sec:MAM}

As described in the previous section, in pure anomaly-mediated SUSY
breaking, soft terms are determined by RG-invariant expressions
involving the gauge and Yukawa couplings.  The soft terms are
therefore completely fixed by the low energy values of these couplings
and an overall scale $\maux$.  If a scalar has negligible Yukawa
interactions, its squared mass is determined by gauge coupling
contributions $- \sum_i b_i g_i^4$, where the sum is over all gauge
groups under which the scalar is charged, and (positive) constants
have been omitted (see Appendix).  {}From this form, we see that
sleptons, which interact only with non-asymptotically free groups
($b_i>0$), have negative squared masses.  Tachyonic sleptons are the
most glaring problem of the anomaly-mediated scenario.

Several mechanisms for solving the tachyonic slepton problem have been
proposed.  Additional positive contributions to slepton squared masses
may arise from bulk contributions~\cite{RS}, gauge-mediated-like
contributions~\cite{PR}, new Yukawa interactions~\cite{Luty}, or
non-decoupling higher order threshold effects~\cite{KSS}.  Here, we
adopt a simple phenomenological approach~\cite{GGW}: we assume an
additional, universal, non-anomaly-mediated contribution $m_0^2$ to
all scalars at the grand unified theory (GUT) scale $\mgut$.  The
resulting boundary conditions,

\begin{eqnarray}
M_\lambda (M_{\rm GUT}) &=& M_\lambda |_{\rm AM} (M_{\rm GUT}) \\
m^2 (M_{\rm GUT}) &=& m^2|_{\rm AM} (M_{\rm GUT}) + m_0^2
\label{bc} \nonumber \\
A (M_{\rm GUT}) &=& A|_{\rm AM} (M_{\rm GUT})\ , \nonumber
\end{eqnarray}
define the minimal anomaly-mediated model.  For large enough $m_0^2$,
slepton squared masses are therefore positive, and the tachyonic
slepton problem is averted.  Such a universal term may be produced by
bulk interactions~\cite{RS}, but is certainly not a feature common to
all anomaly-mediated scenarios.  The extent to which the following
results depend on this assumption will be addressed in
Sec.~\ref{sec:conclusions}.

The addition of a non-anomaly-mediated term destroys the feature of RG
invariance.  However, the RG evolution of the resulting model
nevertheless inherits some of the simplicity of the original pure
anomaly-mediated relations.  Schematically, scalar masses $m_i$
satisfy the one-loop RG equations

\begin{equation}
\frac{d}{dt} m^2_i \sim \frac{1}{16\pi^2} \left[
- g^2 M_{\lambda}^2 + A^2 + \sum_{j} Y^2 m_j^2 \right] \ ,
\label{rge}
\end{equation}
where $t\equiv \ln (\mu/\mgut)$, positive numerical coefficients have
been omitted, and the sum is over all chiral fields $\phi_j$
interacting with $\phi_i$ through the Yukawa coupling $Y$. Letting
$m_i^2 \equiv m_i^2 |_{\text{AM}} + \delta m_i^2$, where $m_i^2
|_{\text{AM}}$ is the pure anomaly-mediated value, the RG invariance
of the anomaly-mediated masses implies

\begin{equation}
\frac{d}{dt} \delta m^2_i \sim \frac{1}{16\pi^2} \sum_j Y^2
\delta m_j^2 \ .
\end{equation}
Thus, at one-loop, the deviations from pure anomaly-mediated relations
satisfy simple evolution equations that depend only on the deviations
themselves.  For scalars with negligible Yukawa couplings, such as the
first and second generation squarks and sleptons, the deviation
$\delta m^2_i$ is a constant of RG evolution.  For them, $\delta
m^2_i$ is simply an additive constant, and the weak scale result for
$m^2_i$ is independent of the scale at which $\delta m_i^2$ is
generated.  For fields interacting through large Yukawa couplings such
as the top Yukawa coupling, the deviations $\delta m_i^2$ evolve;
however, this evolution is simply analyzed.  We will see an important
consequence of this evolution for naturalness in
Sec.~\ref{sec:naturalness}.

We will assume that the boundary conditions of Eq.~(\ref{bc}) are
given at $M_{\rm GUT}=2\times 10^{16}~{\rm GeV}$.  The SUSY breaking
parameters are then evolved with one-loop RG equations to the
superparticle mass scale $m_{\rm SUSY}$, which we have approximated to
be the squark mass scale.  For the gaugino mass parameters, we also
include the largest next-to-leading order corrections from $\alpha_s$
and $\alpha_t \equiv y_t^2/ 4\pi$ given in Ref.~\cite{GGW}.

All parameters of the theory are then specified, except for $\mu$, the
Higgsino mass parameter, and $B_\mu$, the soft bilinear Higgs
coupling.  We do not specify the mechanism for generating these
parameters, but assume that they are constrained so that electroweak
symmetry is properly broken.  Given the other soft parameters at
$m_{\rm SUSY}$, the Higgs potential is determined by $\mu$ and
$B_\mu$, or alternatively, by the Fermi constant $G_{\rm F} =
[2\sqrt{2}(\langle H_u^0\rangle^2 + \langle H_d^0 \rangle^2)]^{-1}
\simeq 1.17\times 10^{-5}~{\rm GeV}^{-2} $ (or, equivalently, the $Z$
mass) and $\tb = \langle H_u^0 \rangle / \langle H_d^0 \rangle$.  It
is more convenient to use the latter two as inputs; $\mu$ and $B_\mu$
are then fixed so that the Higgs potential has a proper minimum with
correct $G_{\rm F}$ and $\tb$.  We minimize the Higgs potential at
one-loop, including radiative corrections from third generation quarks
and squarks~\cite{PRD45-2482}, but neglecting radiative corrections
from other particles.

In fact, the constraint of proper electroweak symmetry breaking does
not determine the sign of the $\mu$ parameter.\footnote{In general,
$\mu$ is a complex parameter, and its phase cannot be determined from
the radiative breaking condition.  In Sec.~\ref{sec:edm}, we consider
the implications of complex $\mu$.  However, in the rest of the paper,
we assume that $\mu$ is real.  In the anomaly-mediated framework,
there are several models in which all CP-violating phases in SUSY
parameters are absent~\cite{RS,PR,KSS}.}  The entire parameter space
of the minimal anomaly-mediated model is therefore specified by 3+1
parameters:

\begin{equation} 
\maux, m_0, \tb, \ \text{and} \ \sign(\mu) \ .  
\end{equation}

\section{Naturalness}
\label{sec:naturalness}

Supersymmetric theories are considered natural from the point of view
of the gauge hierarchy problem if the electroweak scale is not
unusually sensitive to small variations in the underlying parameters.
There are a variety of prescriptions for quantifying naturalness with
varying degrees of sophistication~\cite{AC}.  For the present
purposes, we simply consider a set of parameters to be natural if no
large cancellations occur in the determination of the electroweak
scale.  At tree-level, the relevant condition is

\begin{equation}
\frac{1}{2} m_Z^2 = \frac{m_{H_d}^2 - m_{H_u}^2 \tan^2\beta }
{\tan^2\beta -1} - \mu^2 \ ,
\label{naturalness}
\end{equation}
where $m_{H_u}^2$ and $m_{H_d}^2$ are the soft SUSY breaking masses
for up- and down-type scalar Higgses.  Naturalness then requires that
$|\mu|$ as determined from electroweak symmetry breaking not be too
far above the electroweak scale.  A typical requirement is $|\mu| \alt
1\tev$.

\begin{figure}[tbp]
\postscript{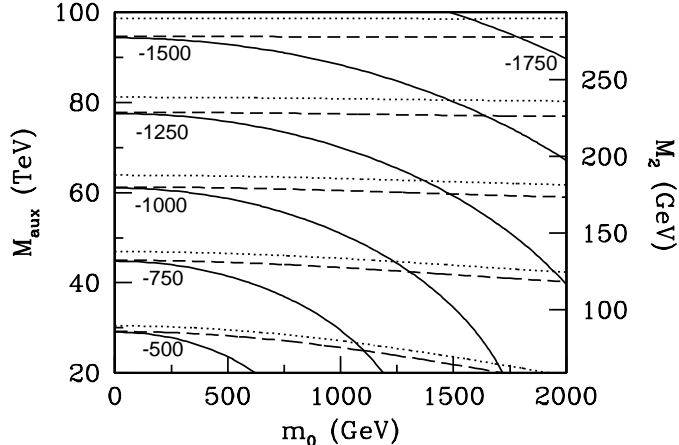}{0.54}
\caption{Contours of constant $\mu<0$ in GeV for $\tb = 3$ (solid), 10
(dashed), and 30 (dotted). The value of $|\mu|$ is similar for
$\mu>0$.}
\label{fig:mu}
\end{figure}

In Fig.~\ref{fig:mu} we present values of $\mu$ in the $(m_0, \maux)$
plane for three representative values of $\tb$: 3 (low), 10
(moderate), and 30 (high).  We have chosen $\mu < 0$ to avoid
constraints from $\bsg$ at large $\tb$ (see Sec.~\ref{sec:bsg}), but
similar $|\mu|$ are found for $\mu>0$.  The parameter $\maux$ is not
phenomenologically transparent, and so on the right-hand axis, we also
give approximate values of the Wino mass $M_2$, using $M_2 =
\frac{g_2^2}{16 \pi^2} \maux \simeq 2.9 \times 10^{-3}\maux$.

The value of $|\mu|$ rises with increasing $\maux$, as expected.
Irrespective of $m_0$ and $\tb$, $|\mu| \alt 1\tev$ implies $M_2 \alt
200 \gev$.  Such a restriction is encouraging for searches for
degenerate Winos at upcoming runs of the Tevatron, as will be
discussed more fully in Sec.~\ref{sec:charginos}.

The special case of $m_0=0$ corresponds to pure anomaly-mediated SUSY
breaking.  In this case, the expressions for soft SUSY breaking terms
are RG-invariant and the soft masses may be evaluated at any scale,
including a low (TeV) scale.  Based on this observation, it has been
argued that, since the stop masses do not enter the determination of
$m_{H_{u,d}}^2$ with large logarithms through RG evolution, stop
masses of $2\tev$ or even higher are consistent with
naturalness~\cite{RS}.  This is contradicted by Fig.~\ref{fig:mu}: for
$m_0=0$, as will be seen in Sec.~\ref{sec:squarks}, stop masses of
$2\tev$ require very large $\maux$ corresponding to values of $|\mu|$
above 2 TeV.  Stop masses of 2 TeV are therefore as unnatural in {\em
pure} anomaly-mediated SUSY breaking as they are in more conventional
gravity-mediated scenarios, such as minimal supergravity.  This
applies to all cases where the pure anomaly-mediated relations are
approximately valid for squark and Higgs soft masses, and includes
models in which a mechanism for avoiding tachyonic sleptons is invoked
which does not disturb the squark and Higgs masses.

For the minimal anomaly-mediated model with $m_0>0$, however, the
squark and Higgs masses are explicitly modified, and the argument
above does not apply.  It is exactly in this case, where the soft SUSY
masses are {\em not} RG-invariant, that there is the possibility that
heavy squarks can be consistent with naturalness, and we will see
that, in fact, this is realized by a novel mechanism for large $m_0$.

In Fig.~\ref{fig:mu}, for $\tb=3$, an upper bound on $|\mu|$ implies
an upper bound on $m_0$.  However, for moderate and large $\tb$, the
contours of constant $|\mu|$ are extremely insensitive to $m_0$, and
so large squark and slepton masses are consistent with naturalness in
the large $m_0$ regime.\footnote{In Ref.~\cite{GGW}, the insensitivity
of $|\mu|$ to $m_0$ is implicit in Fig.~1; its implications for
naturalness were not noted.}  This behavior may be understood first by
noting that, for moderate and large $\tb$, Eq.~(\ref{naturalness})
implies that $\mu$ depends sensitively on $m_{H_u}^2$ only.  The RG
evolution of $m_{H_u}^2$ is most easily understood by letting
$m_{H_u}^2 \equiv m_{H_u}^2 |_{\text{AM}} + \delta m_{H_u}^2$, where
$m_{H_u}^2 |_{\text{AM}}$ is the pure anomaly-mediated value, and
similarly for all other scalar masses.  The deviations $\delta m_i^2$
satisfy simple RG equations, as discussed in Sec.~\ref{sec:MAM}.  For
$\tb$ not extremely large, the only large Yukawa is the top Yukawa
$Y_t$, and $m_{H_u}^2$ is determined by the system of RG equations

\begin{equation}
\frac{d}{dt} \left( \begin{array}{c} \delta m_{H_u}^2
\\ \delta m_{U_3}^2 \\ \delta m_{Q_3}^2 \end{array} \right)
= \frac{Y_t^2}{8\pi^2} \left(
\begin{array}{ccc}
3 & 3 & 3 \\
2 & 2 & 2 \\
1 & 1 & 1 \end{array} \right)
\left( \begin{array}{c} \delta m_{H_u}^2
\\ \delta m_{U_3}^2 \\ \delta m_{Q_3}^2 \end{array} \right) \ ,
\end{equation}
where $Q_3$ and $U_3$ denote the third generation squark SU(2) doublet
and up-type singlet representations, respectively.

Such systems of RG equations are easily solved by decomposing
arbitrary initial conditions into components parallel to the
eigenvectors of the evolution matrix, which then evolve
independently~\cite{baggerwise}. In the present case, the solution
with initial condition $m_0^2 (1,1,1)^T$ is

\begin{equation}
\left( \begin{array}{c} \delta m_{H_u}^2
\\ \delta m_{U_3}^2 \\ \delta m_{Q_3}^2 \end{array} \right)
= \frac{m_0^2}{2}
\left( \begin{array}{c} 3 \\ 2 \\ 1 \end{array} \right)
{\rm exp} \left[ 6 \int_0^t \frac{Y_t^2}{8\pi^2} dt' \right]
-\frac{m_0^2}{2}
\left( \begin{array}{c} 1 \\ 0 \\ -1 \end{array} \right) \ .
\end{equation}
For $t$ and $Y_t$ such that ${\rm exp} \left[ 6 \int_0^t
\frac{Y_t^2}{8\pi^2} dt' \right] = 1/3$, $\delta m_{H_u}^2 = 0$, i.e.,
$m_{H_u}^2$ assumes its pure anomaly-mediated value for any $m_0$.

\begin{figure}[tb]
\postscript{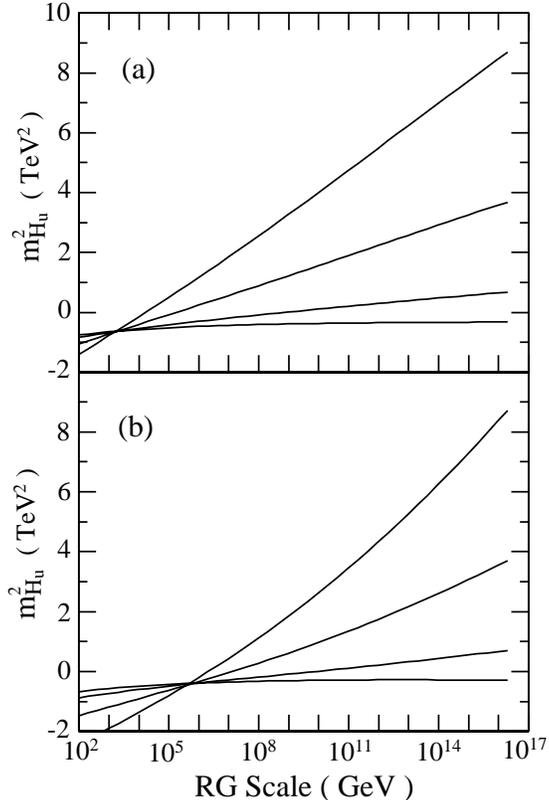}{0.44}
\caption{The RG evolution of $m_{H_u}^2$ for fixed $\maux=50 \tev$,
$\tb=10$, and top quark mass (a) 174 GeV (b) 184 GeV.  The GUT scale
boundary conditions are for $m_0 = 0, 1, 2, 3 \tev$, from the bottom.
The RG behavior of $m_{H_u}^2$ exhibits a focus point (not a fixed
point) near the weak scale, where $m_{H_u}^2$ takes its pure
anomaly-mediated value, irrespective of $m_0$. }
\label{fig:run}
\end{figure}

The RG evolution of $m_{H_u}^2$ is shown for several values of $m_0$
in Fig.~\ref{fig:run}.  As expected, the RG curves intersect at a
single point where $m_{H_u}^2$ is independent of $m_0$; we will call
this a ``focus point.''  Remarkably, however, the focus point occurs
near the weak scale for $Y_t$ corresponding to the physical top mass
of $\mt \approx 174 \gev$.  Thus the weak scale value of $m_{H_u}^2$
is nearly its pure anomaly-mediated value for all values of
$m_0$. Note that this behavior applies only to $m_{H_u}^2$; no other
scalar mass has a focus point behavior.

\begin{figure}[tbp]
\postscript{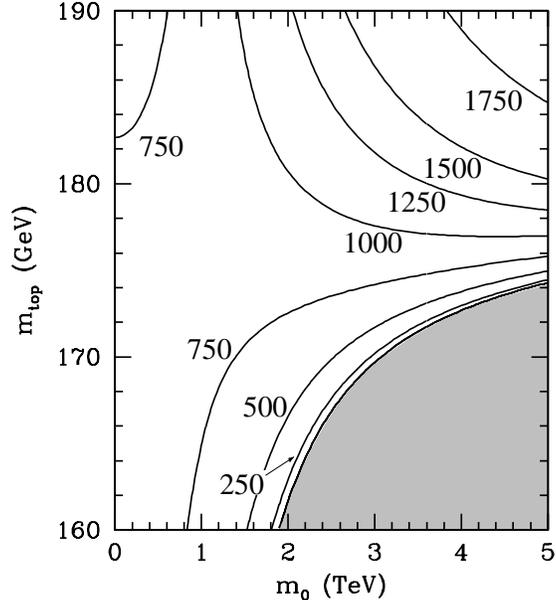}{0.44}
\caption{Contours of $|\mu|$ in GeV for $\mu<0$, $\maux=50 \tev$, and
$\tb=10$.  In the shaded region, electroweak symmetry cannot be broken
radiatively. }
\label{fig:naturalness}
\end{figure}

The focus point is not a fixed point; for example, below the focus
point, the RG curves diverge again.  The position of the focus point
depends on $Y_t$, and we must check the sensitivity to variations in
$Y_t$.  In Fig.~\ref{fig:run} we show also the behavior for $Y_t$
corresponding to $\mt= 184$ GeV.  The exact weak scale value of
$m_{H_u}^2$ depends on $Y_t$ and, when the focus point is not exactly
at the weak scale, also on $m_0$.  However, for top quark masses near
the physical one, the focus point remains within a couple of decades
of the weak scale, and the sensitivity to variations in $m_0$ is
always suppressed.  This is demonstrated in
Fig.~\ref{fig:naturalness}, where values of $\mu$ are given in the
$(m_0, \mt)$ plane.  Even for $m_0^2 = 25 \tev^2$ and $\mt = 174 \pm
5\gev$, we find that $\mu^2$ lies naturally below 2 $\tev^2$.

An interesting question is whether $m_0$ can be bigger than the weak
scale by a loop factor without compromising naturalness.  If this were
the case, there would be no need to appeal to a sequestered sector to
eliminate tree-level scalar masses.  However, $m_0$ cannot be
arbitrarily large.  In Fig.~\ref{fig:naturalness}, we see that the
requirement of proper electroweak symmetry breaking implies $m_0 \alt
5\tev$.  In any case, a similar bound would follow from requiring that
one-loop finite corrections to the Higgs squared mass parameter, which
are proportional to $m_{\tilde{f}}^2$, not introduce large
fine-tunings. The maximum allowed $m_0^2$ is thus roughly an order of
magnitude below $\maux^2$.  Thus, while it is possible to eliminate
the sequestered seor mechanism for direct K\"ahler interaction
suppression, it is still required that the tree-level scalar squared
mass $m_0^2$ be suppressed by an order of magnitude relative to its
``natural'' value $\sim \maux^2$.

Nevertheless, given that we have no understanding of the source of
$m_0$, it is at least somewhat reassuring that it may be far above the
weak scale without incurring a fine-tuning penalty.  A direct
consequence of this is that the minimal anomaly-mediated model is a
model that naturally accommodates multi-TeV sleptons and squarks.  As
we will see below, this has important phenomenological consequences
both for high energy colliders and low energy probes.

\section{Superpartner Spectra and Implications for High Energy
Colliders}
\label{sec:highenergy}

Having defined the minimal anomaly-mediated model in
Sec.~\ref{sec:MAM} and explored the natural range of its fundamental
parameters in Sec.~\ref{sec:naturalness}, we now consider the
resulting masses and mixings of the superpartners.  The lightest
supersymmetric particles are either a degenerate triplet of charginos
and neutralinos, the lighter stau $\stau_1$, or the tau sneutrino
$\tilde{\nu}_{\tau}$. We begin by considering these, and conclude with
a discussion of the squark spectrum. We do not discuss the gluino and
heavy Higgses in detail.  However, their masses are given in
Eq.~(\ref{gauginoratios}) and Figs.~\ref{fig:bsgp} and \ref{fig:bsgn},
respectively.

\subsection{Charginos and Neutralinos}
\label{sec:charginos}

\begin{figure}[tbp]
\postscript{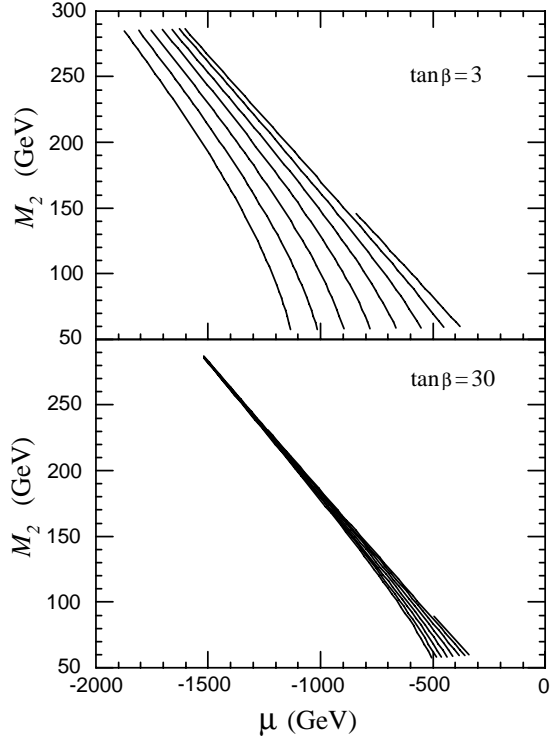}{0.44}
\caption{Parameters allowed by current constraints on particle masses
and radiative symmetry breaking in the $(\mu, M_2)$ plane for $\mu <
0$ and $\tb=3$ and 30. The contours are for constant $m_0 = 250$, 500,
$\cdots$, 2000 GeV from the right.  Similar results hold for $\mu >
0$. }
\label{fig:mum2}
\end{figure}

Charginos and neutralinos are mixtures of gauginos and Higgsinos.
Their composition is determined by $M_2$, $M_1$, $\mu$, and $\tb$ at
tree-level.  Inserting the values of the gauge coupling constants at
$m_Z$ in Eq.~(\ref{M_G(AMSB)}), and including the largest
next-to-leading corrections as described in Sec.~\ref{sec:MAM}, we
find

\begin{equation}
M_1 : M_2 :  M_3 \approx 2.8 : 1 : -8.3 \ .
\label{gauginoratios}
\end{equation}
Typical values of $(\mu, M_2)$ allowed by radiative electroweak
symmetry breaking in the minimal anomaly-mediated model are given in
Fig.~\ref{fig:mum2}.  Combined with the anomaly-mediated relation $M_1
\simeq 2.8 M_2$, Fig.~\ref{fig:mum2} implies $M_2 < M_1 < |\mu|$ with
substantial hierarchies in these parameters throughout parameter
space.  The chargino and neutralino mass eigenstates are therefore
well-approximated by pure gaugino and pure Higgsino states with masses

\begin{eqnarray}
M_2: && \neutralino_1 \simeq \tilde{W}^0 \ , \
\chargino_1 \simeq \tilde{W}^{\pm} \nonumber \\
M_1: && \neutralino_2 \simeq \tilde{B} \\
|\mu|: && \neutralino_{3,4} \simeq \tilde{H}_u^0 \pm \tilde{H}_d^0\ ,
\ \chargino_2 \simeq \tilde{H}^{\pm} \ , \nonumber
\end{eqnarray}
and the lightest of these is always a highly degenerate triplet of
Winos.

In much of parameter space, as we will see in Sec.~\ref{sec:sleptons},
these Winos are the LSPs.  The possibility of searching for
supersymmetry in the Wino LSP scenario has been the subject of much
recent attention~\cite{FMRSS,GGW,CDG,CDM,GM}.  The detection of Wino
LSPs poses novel experimental challenges.  Neutral Winos pass through
collider detectors without interacting.  Charged Winos are detectable
in principle, but are typically highly degenerate with neutral Winos,
with $\Delta m = m_{\chargino_1} - m_{\neutralino_1} \approx 150 - 300
\mev$ and corresponding decay lengths $c\tau =0.5 -10$
cm~\cite{FMRSS,GGW,CDG,CDM,GM,TW}.  They therefore decay to invisible
neutral Winos and extremely soft pions before reaching the muon
chambers, thereby escaping both conventional searches based on
energetic decay products and searches for long-lived charged particles
that produce hits in the muon chamber.

Fig.~\ref{fig:mum2}, however, has two important and encouraging
implications for Wino LSP searches.  First, as noted in
Sec.~\ref{sec:naturalness}, naturalness bounds on $|\mu|$ imply
stringent bounds on $M_2$.  {}From Fig.~\ref{fig:mum2}, for example,
we find that $|\mu| \alt 1 \tev$ implies $M_2 \alt 200 \gev$.
Continuing searches at LEP~\cite{delphi}, although limited
kinematically to the region $M_2 \alt 100 \gev$, will be able to probe
a significant fraction of this parameter region.  In addition, such
limits on the Wino mass imply large cross sections at the Tevatron.
For $M_2 = 200 \gev$ and $\sqrt{s} = 2 \tev$, the Wino pair production
rate is $\sigma ( p\bar{p} \to \tilde{W}^\pm \tilde{W}^0,
\tilde{W}^\pm \tilde{W}^\mp ) \approx 100$ fb, and if a jet with $p_T
> 30\gev$ and $|\eta| < 2$ is required for triggering, the associated
production rate is $\sigma ( p\bar{p} \to \tilde{W}^\pm \tilde{W}^0 +
\text{ jet}, \tilde{W}^\pm \tilde{W}^\mp+ \text{ jet}) \approx 10$
fb~\cite{FMRSS}.  Such cross sections imply hundreds of Wino pairs
produced at the upcoming Run II, and tens of Wino pairs produced in
association with jets.

Second, the region of $(\mu, M_2)$ space favored in
Fig.~\ref{fig:mum2} is the far gaugino region, where $\Delta m$ is
minimized. For the parameters of Fig.~\ref{fig:mum2}, $\Delta m <
180\mev$, corresponding to decay lengths of $c\tau > 3.5$ cm. (See
Ref.~\cite{FMRSS}.)  Thus, a significant fraction of Winos will pass
through several vertex detector layers. When produced in association
with a jet for triggering, such Winos will be discovered off-line as
high $dE/dx$ tracks with no associated calorimeter or muon chamber
activity.  Such a signal should be spectacular and background-free.
This possibility is discussed in detail in Ref.~\cite{FMRSS}, where an
integrated luminosity of 2 $\ifb$ is shown to probe the entire region
discussed here with $|\mu| < 1\tev$.  It is exciting that Run II of
the Tevatron will either discover Wino LSPs or exclude most of the
natural region of parameter space in this model.

\subsection{Sleptons}
\label{sec:sleptons}

Slepton masses and mixings are given by the mass matrix

\begin{equation}
\bold{M_{\tilde{l}}^2} =
\left( \begin{array}{cc}
m_{\tilde{L}}^2 + m_l^2 - m_Z^2(\frac{1}{2} - \sin^2\theta_W)
\cos 2\beta & m_l (A_l - \mu \tb) \\
m_l (A_l - \mu \tb) &
m_{\tilde{E}}^2 + m_l^2 - m_Z^2\sin^2\theta_W\cos 2\beta
\end{array}
\right)
\end{equation}
in the basis $(\tilde{l}_L, \tilde{l}_R)$, and sneutrino masses are
given by

\begin{equation}
m_{\tilde{\nu}}^2 =
m_{\tilde{L}}^2 + \frac{1}{2} m_Z^2 \cos 2\beta \ ,
\end{equation}
where $m_{\tilde{L}}^2$ and $m_{\tilde{E}}^2$ are the soft SUSY
breaking masses.

In anomaly-mediated models, as discussed in Ref.~\cite{GGW}, if both
$m_{\tilde{L}}^2$ and $m_{\tilde{E}}^2$ receive the same $m_0^2$
contribution, the diagonal entries of the slepton mass matrix are
accidentally highly degenerate.  The anomaly-mediated boundary
conditions imply (see the Appendix)

\begin{equation}
\bold{M_{\tilde{l}}^2}_{LL} - \bold{M_{\tilde{l}}^2}_{RR} =
\frac{3}{2} \left( \frac{g_2^2 \maux} {16 \pi^2} \right)^2
\left[ 11\tan^4\theta_W -1 \right] + m_Z^2 \left[2\sin^2\theta_W
- \frac{1}{2} \right] \cos 2\beta \ .
\label{sleptondiff}
\end{equation}
For $\sin^2\theta_W = 0.2312$, $\tan^4\theta_W = 0.0904$, and both
bracketed expressions are extremely small.  This accidental degeneracy
implies that same-flavor sleptons may be highly degenerate.  The
physical mass splitting for staus is given in
Fig.~\ref{fig:staudm}. For low $\tb$ (and, by implication, for all
$\tb$ for selectrons and smuons), degeneracies of order 10 GeV or less
are found throughout the parameter region.  For large $\tb$, however,
large Yukawa effects dilute the degeneracy significantly.

\begin{figure}[tbp]
\postscript{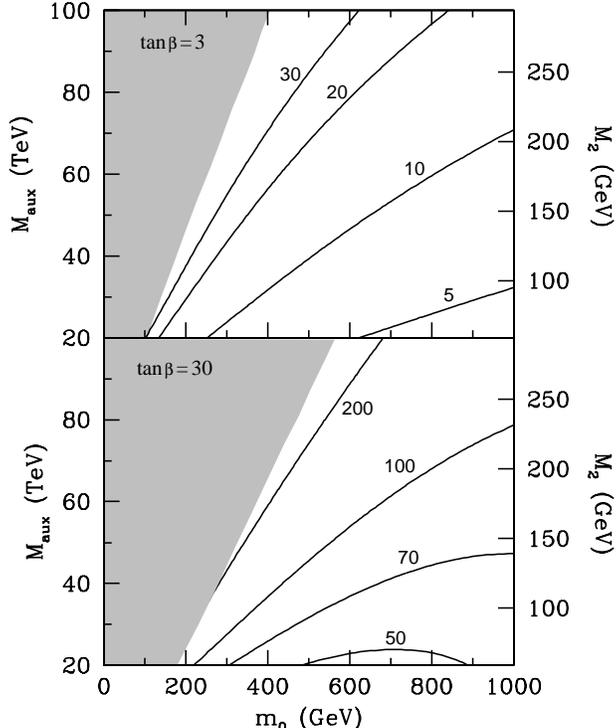}{0.49}
\caption{Contours of constant $m_{\tilde{\tau}_2} -
m_{\tilde{\tau}_1}$ in GeV for $\mu < 0$ and $\tb=3$ and 30. The
shaded region is excluded by $m_{\stau_1}> 70\gev$.}
\label{fig:staudm}
\end{figure}

Equation (\ref{sleptondiff}) also implies that even small off-diagonal
entries may lead to large mixing.  The left-right mixing for staus is
given in Fig.~\ref{fig:stauLR}.  Throughout parameter space, and even
for low $\tb$, the stau mixing is nearly maximal.  In fact, even smuon
mixing may be significant --- for large $\tb$ and low $\maux$, it too
is almost maximal.  Nearly degenerate and highly-mixed same flavor
sleptons are a distinctive feature of the minimal anomaly-mediated
model and distinguish it from other gravity- and gauge-mediated
models, where, typically, $m_{\tilde{l}_L} > m_{\tilde{l}_R}$.  These
features may be precisely tested by measurements of slepton
masses~\cite{sleptonmasses} and mixings~\cite{stau} at future
colliders.

\begin{figure}[tbp]
\postscript{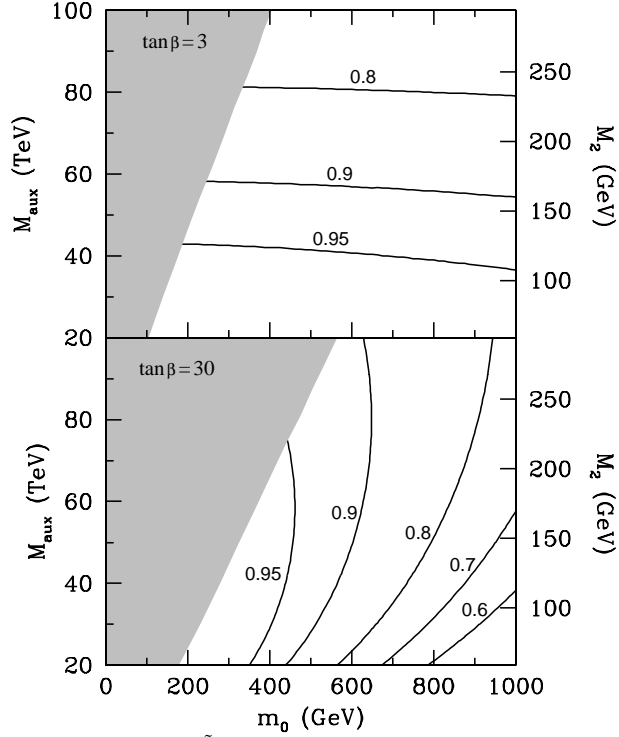}{0.49}
\caption{Contours of constant $|\sin 2 \theta^{\stau}_{LR}|$ for $\mu
< 0$ and $\tb=3$ and 30. The shaded region is excluded by
$m_{\stau_1}> 70\gev$.}
\label{fig:stauLR}
\end{figure}

The lighter stau $\stau_1$ is always the lightest charged slepton, and
it therefore plays an important phenomenological role.  The $\stau_1$
mass is displayed in Fig.~\ref{fig:staum}. 
\begin{figure}[tbp]
\postscript{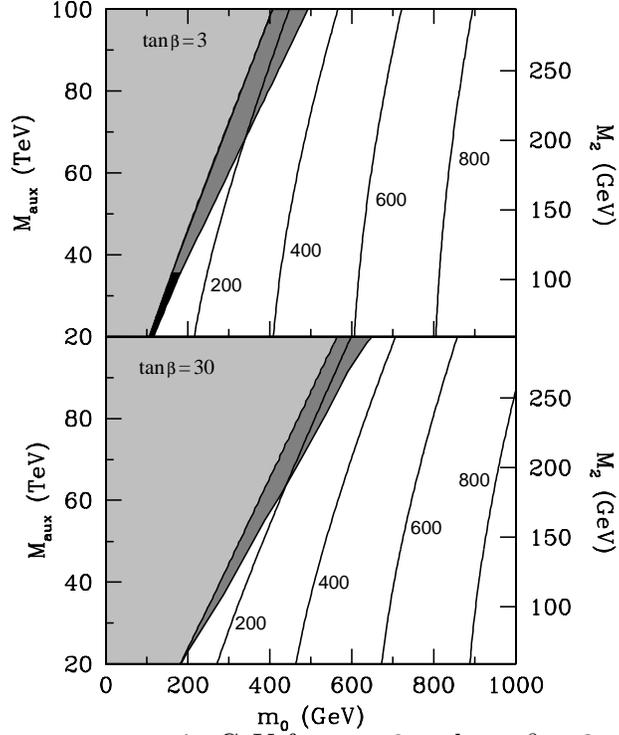}{0.49}
\caption{Contours of constant $m_{{\stau}_1}$ in GeV for $\mu < 0$ and
$\tb=3$ and 30. The light shaded region is excluded by $m_{\stau_1}>
70\gev$. In the dark shaded region, the LSP is $\stau_1$, in the
blackened region, the LSP is $\tilde{\nu}_{\tau}$, and in the unshaded
region, the LSPs are a highly degenerate triplet of Winos.}
\label{fig:staum}
\end{figure}
For low $m_0$, $\stau_1$ is either tachyonic or excluded by
experimental bounds. The current bounds are fairly complicated in this
model, since the mass ordering and mass splittings between $\stau_1$,
the Winos, and the sneutrinos vary throughout the parameter space.
For staus decaying to neutralinos with a mass splitting greater than
15 GeV, combined LEP analyses of the $\sqrt{s} = 189\gev$ data yield
the bound $m_{\stau} > 71 \gev$~\cite{LEPwebpage}, but this drops to
near the LEP I limit of 45 GeV as the mass splitting goes to zero.
However, for stable staus, combined LEP analyses of data up to
$\sqrt{s}=183\gev$ imply $m_{\stau} > 87\gev$~\cite{LEPwebpage}.  The
light shaded region of Fig.~\ref{fig:staum} is excluded by $m_{\stau}
> 70\gev$ and represents a rough summary of these bounds. In the
remaining region, the bounds $m_{\tilde{\nu}} > 43
\gev$~\cite{snubound}, $m_{\tilde{e}} > 89\gev$~\cite{LEPwebpage}, and
$m_{\tilde{\mu}} > 84\gev$~\cite{LEPwebpage} are always satisfied.  In
the following, we will include the excluded shaded region in plots of
observables that involve sleptons.  For quantities such as squark
masses or rates for $\bsg$, we omit this, as such quantities are
well-defined even for small $m_0$, and in fact, the $m_0=0$ axis gives
their values in anomaly-mediated models where the slepton mass problem
is fixed without changing the squark and Higgs masses.

For large $m_0$, $m_{\stau_1}\approx m_0$, and the Wino is the LSP.
This is the case in the unshaded region of Fig.~\ref{fig:staum}.  The
experimental implications of the Wino LSP scenario have been discussed
above in Sec.~\ref{sec:charginos}.

Finally, there exists an intermediate $m_0$ region, in which the LSP
is either the $\stau_1$ or the $\tilde{\nu}_\tau$.  In the $\stau_1$
LSP scenario (the dark shaded region of Fig.~\ref{fig:staum}), the
stau may be found at both LEP and the Tevatron through its spectacular
anomalous $dE/dx$ and time-of-flight
signatures~\cite{LEPwebpage,stauth,stauexp}.  At the Tevatron, for
example, for $m_{\stau_1}\alt 150 \gev$, $\sigma (p\bar{p} \to \stau_1
\stau_1^*) \agt 1$ fb, and so a significant fraction of the stau LSP
parameter space may be explored.\footnote{Note that in this parameter
region, the stau is absolutely stable, assuming R-parity
conservation. (Recall that the gravitino mass is of order $\maux$.)
This scenario therefore requires some mechanism for diluting the stau
density, such as late inflation with a low reheating
temperature~\cite{lateinflation}.}

In the case of the sneutrino LSP (the blackened region of
Fig.~\ref{fig:staum}), there are many possible experimental
signatures.  While this region appears only for a limited range of
SUSY parameters, superparticles tend to be relatively light in this
region, with $m_{\stau_1} \alt 100 \gev$ and $M_2 \alt 110\gev$, and
so it is amenable to study at LEP.  In this region, the slepton mass
ordering is always

\begin{equation}
\tilde{\nu}_{\tau}, \tilde{\nu}_{\mu}, \tilde{\nu}_e
< \stau_1 < \tilde{e}_R, \tilde{\mu}_1
< \tilde{e}_L, \tilde{\mu}_2 < \stau_2 \ ,
\end{equation}
and the Wino triplet may appear anywhere between the sneutrinos and
$\stau_2$.  Typically, though not always, the only kinematically
accessible superparticles at LEP are the sneutrinos, $\stau_1$ and the
Winos.  The two possible mass orderings and dominant decay modes in
each scenario are then

\begin{eqnarray}
\stau_1 > \tilde{W}^{\pm,0} > \tilde{\nu} :\ &&
\stau_1 \to \tau \tilde{W}^0, \nu_{\tau} \tilde{W}^{\pm} \nonumber \\
&&\tilde{W}^0 \to \nu_l \tilde{\nu}_l \ , \
\tilde{W}^\pm \to l \tilde{\nu}_l \\
\tilde{W}^{\pm,0} > \stau_1 > \tilde{\nu} :\ &&
\tilde{W}^0 \to \nu \tilde{\nu} , \tau \stau_1 \ , \
\tilde{W}^\pm \to l \tilde{\nu}_l , \nu_{\tau} \stau_1 \nonumber \\
&& \stau_1 \to \pi^{\pm} \tilde{\nu}_{\tau} \ .
\end{eqnarray}

\subsection{Squarks}
\label{sec:squarks}

In anomaly-mediated SUSY breaking, squarks are universally very heavy,
as their masses receive contributions from the strong coupling.  The
gauge coupling contribution to scalar squared masses is of the form
$-b_i g_i^4$, where $b_i$ is the one-loop $\beta$-function coefficient
(see Appendix), and so the strong coupling contribution completely
overwhelms those of the SU(2) and U(1) couplings.  Squark masses for
the first two generations are therefore both flavor- and
chirality-blind; we find that the $\tilde{u}_L$, $\tilde{u}_R$,
$\tilde{d}_L$, and $\tilde{d}_R$, and their second generation
counterparts are all degenerate to within $\sim 10$ GeV throughout
parameter space.

The first and second generation squark masses are given in
Fig.~\ref{fig:sq}.  The squarks are hierarchically heavier than Winos
and sleptons for low $m_0$, and their mass increases as $m_0$
increases.  For $m_0 \agt 2\tev$, the squark mass is above 2
TeV. Thus, the focus point naturalness behavior discussed in
Sec.~\ref{sec:naturalness}, which allows such large $m_0$, has
important phenomenological consequences.  Direct detection of 2 TeV
squarks is likely to be impossible at the LHC or NLC, and must wait
for even higher energy hadron or muon colliders. Note, however, that
some superparticles, notably the gauginos, cannot evade detection at
the LHC and NLC.

\begin{figure}[tbp]
\postscript{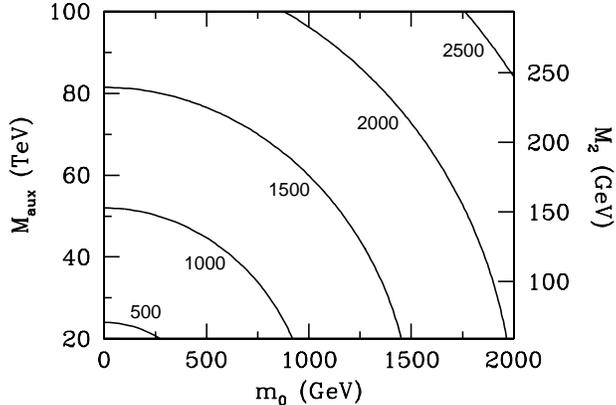}{0.49}
\caption{Contours of constant $m_{\tilde{u}_L}$ in GeV for $\mu < 0$.
The masses of all first and second generation squarks are degenerate
to within $\sim 10$ GeV throughout the parameter space and are
insensitive to $\tb$.}
\label{fig:sq}
\end{figure}

Unlike the squarks of the first two generations, the masses of third
generation squarks $\tilde{t}_L$, $\tilde{t}_R$, $\tilde{b}_L$, and
(for large $\tb$) $\tilde{b}_R$ receive significant contributions from
large Yukawa couplings.  These are shown in Figs.~\ref{fig:stop} and
\ref{fig:sbot} for small and large values of $\tb$.  Yukawa couplings
always reduce the masses and their effect may be large. For example,
$m_{\tilde{t}_1}$ may be reduced by as much as 40\% relative to the
first and second generation squark masses.  At the LHC, therefore,
stops and sbottoms may be produced in much larger numbers than the
other squarks, adding to the importance of $b$-tagging.

\begin{figure}[tbp]
\postscript{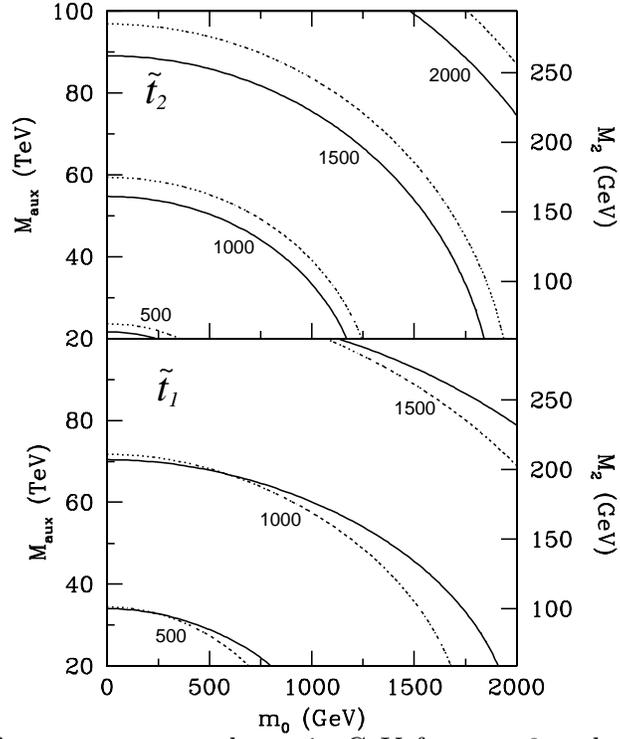}{0.49}
\caption{Contours of constant $m_{\tilde{t}_2}$ and $m_{\tilde{t}_1}$
in GeV for $\mu < 0$ and $\tb = 3$ (solid) and 30 (dotted).}
\label{fig:stop}
\end{figure}

\begin{figure}[tbp]
\postscript{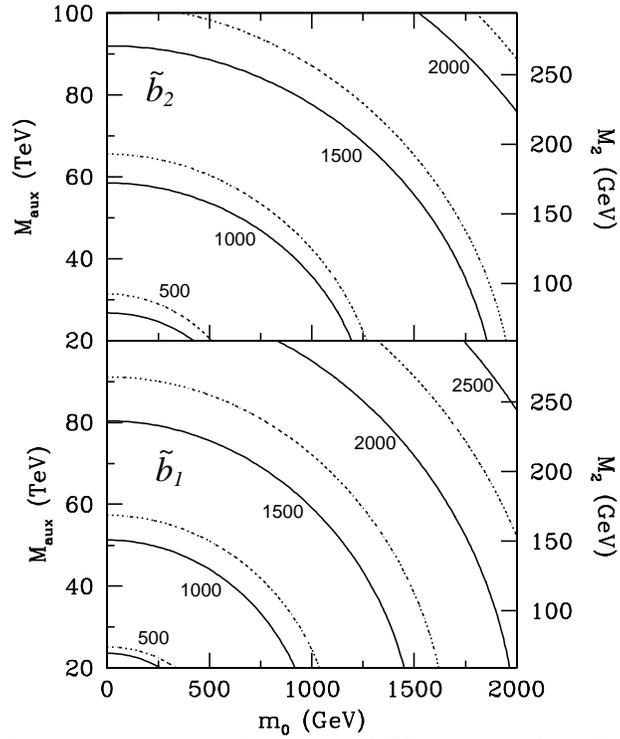}{0.49}
\caption{Contours of constant $m_{\tilde{b}_2}$ and $m_{\tilde{b}_1}$
in GeV for $\mu < 0$ and $\tb = 3$ (solid) and 30 (dotted).}
\label{fig:sbot}
\end{figure}

As in the case of sleptons, third generation squarks may have large
left-right mixing.  For $\tb=30$, left-right mixing in both the stops
and sbottoms is large, and is nearly maximal for low $m_0$.  For
$\tb=3$, sbottom mixing is negligible, but stop mixing may still be as
large as $\sin 2\theta^{\tilde{t}}_{LR} \approx 0.2$.

\section{Low Energy Probes}
\label{sec:lowenergy}

Anomaly-mediated supersymmetry breaking naturally suppresses
flavor-violation in the first and second generations, but not all low
energy constraints are therefore trivially satisfied.  In particular,
since anomaly-mediated soft terms depend on Yukawa couplings,
non-trivial flavor mixing involving third generation squarks can be
expected.  We first study the flavor-changing process $\bsg$, which is
well-known for being sensitive to third generation flavor violation.
We then consider magnetic and electric dipole moments, observables
that are flavor-conserving, but are nevertheless highly sensitive to
SUSY effects.

\subsection{\bold{\bsg}}
\label{sec:bsg}

In the standard model, the flavor-changing transition $\bsg$ is
mediated by a $W$ boson at one-loop.  In supersymmetric theories,
$\bsg$ receives additional one-loop contributions from charged Higgs-,
chargino-, gluino-, and neutralino-mediated processes.  The charged
Higgs contribution depends only on the charged Higgs mass and $\tb$,
interferes constructively with the standard model amplitude, and is
known to be large even for charged Higgs masses beyond current direct
experimental bounds.  The supersymmetric contributions may also be
large for some ranges of SUSY parameters.  Thus, $\bsg$ provides an
important probe of all supersymmetric models, including those that are
typically safe from other flavor-violating constraints.

In the well-studied cases of minimal supergravity and gauge-mediated
SUSY breaking~\cite{bsgSUSY}, the chargino- and, to a lesser extent,
gluino-mediated contributions may be significant for large
$\tb$. Neutralino contributions are always negligible.  For $\mu <0$
(in our conventions), these contributions are constructive and so, for
large $\tb$, positive $\mu$ is favored.

In the present case of anomaly-mediated SUSY breaking, several new
features arise.  First, in contrast to the case of minimal
supergravity and gauge-mediation where squark mixing arises only
through RG evolution, flavor violation in the squark sector is present
even in the boundary conditions (and receives additional contributions
from RG evolution).  More importantly, the signs of the parameter
$A_t$ and the gluino mass $M_3$ are opposite to those of minimal
supergravity and gauge-mediation.  The leading contributions for large
$\tb$ in the mass insertion approximation from charginos and gluinos
are given in Fig.~\ref{fig:feynmanbsg}.  For large $\tb$, the
amplitudes ${\cal A}_{\chargino} \propto \sign (\mu A_t)$ and ${\cal
A}_{\tilde{g}} \propto \sign (\mu M_3)$ are both opposite in sign
relative to their values in minimal supergravity and gauge-mediation.

\begin{figure}[tbp]
\postscript{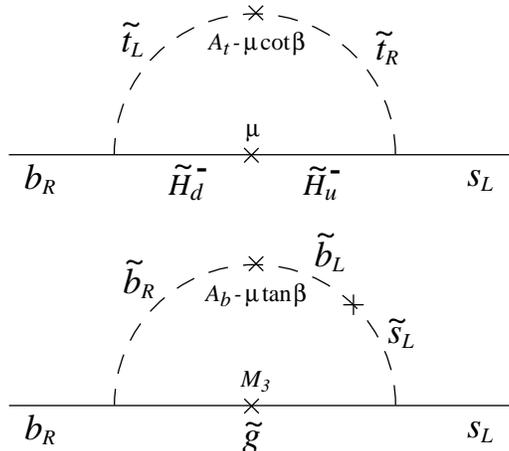}{0.41}
\caption{The leading contributions to $\bsg$ from chargino- and
gluino-mediated processes in the mass insertion approximation for
large $\tb$. The photon, which may couple to any charged internal
propagator, is omitted.}
\label{fig:feynmanbsg}
\end{figure}

$B(B \to X_s \gamma)$ may be calculated by first matching the full
supersymmetric theory on to the effective Hamiltonian

\begin{equation}
{\cal H}_{\text{eff}} = -\frac{4G_F}{\sqrt{2}} V^*_{ts} V_{tb}
\sum_{i=1}^8 C_i {\cal O}_i
\end{equation}
at the electroweak scale $m_W$.  In the basis where the current and
mass eigenstates are identified for $d_L$, $d_R$, and $u_R$,
supersymmetry contributes dominantly to the Wilson coefficients $C_7$
and $C_8$ of the magnetic and chromomagnetic dipole operators

\begin{eqnarray}
{\cal O}_7 &=& \frac{e}{16\pi^2}m_b (\bar{s}_L \sigma^{\mu\nu}
b_R ) F_{\mu\nu} \\
{\cal O}_8 &=& \frac{g_s}{16\pi^2}m_b (\bar{s}_L \sigma^{\mu\nu}
T^a b_R ) G^a_{\mu\nu} \ .
\end{eqnarray}
(Contributions to operators with chirality opposite to those above are
suppressed by $m_s/m_b$ and are negligible.)  We use next-to-leading
order (NLO) matching conditions for the standard model~\cite{NLOSM}
and charged Higgs~\cite{NLOH+} contributions.  The remaining
supersymmetric contributions are included at leading
order~\cite{BBMR}.  Some classes of NLO supersymmetric contributions
have also been calculated~\cite{NLOSUSY}; however, a full NLO
calculation is not yet available.  For the present purposes, where we
will be scanning over SUSY parameter space, the leading order results
are sufficient.  Note that the inclusion of some, but not all, NLO
effects is formally inconsistent, but by doing so, we are effectively
assuming that the NLO corrections in a given renormalization scheme
are numerically small.

The Wilson coefficients $C_i$ at the weak scale are then evolved down
to a low energy scale $\mu_b$ of order $m_b$, where matrix elements
are evaluated using the resulting effective operators.  The NLO
anomalous dimension matrix is now known~\cite{bsgADM}, as are the NLO
matrix elements~\cite{bsgmatrix} and the leading order QED and
electroweak radiative corrections~\cite{bsgQED,KN}.  These have been
incorporated in the analysis of Ref.~\cite{KN}, where a simple form
for $B(B \to X_s \gamma)$ in terms of weak scale Wilson coefficients
is presented.  The exact parametrization depends on the choice of
$\mu_b$ and the photon energy cutoff $E_{\gamma}^{\text{min}} =
\frac{1}{2} (1-\delta) m_B$.  We choose $\mu_b = m_b$ and $\delta =
0.9$.  The SUSY branching fraction is then given by~\cite{KN}

\begin{equation}
\frac{B(B \to X_s \gamma)}{B(B \to X_s \gamma)_{\text{SM}}} =
1 + 0.681 r_7 + 0.116 r_7^2 + 0.0832 r_8 + 0.00455 r_8^2
+ 0.0252 r_7 r_8 \ ,
\end{equation}
where $r_{7,8}$ are the fractional deviations from standard model
amplitudes:

\begin{equation}
r_{7,8} \equiv \frac{C_{7,8}(m_W)}{C_{7,8}^{\text{SM}}(m_W)} - 1
= \left. \frac{{\cal A}_{H^{\pm}} + {\cal A}_{\chargino}
+ {\cal A}_{\tilde{g}} + {\cal A}_{\neutralino}}
{{\cal A}_{\text{SM}}} \right|_{7,8} \ .
\end{equation}
For the standard model value, we take~\cite{KN}

\begin{equation}
B(B \to X_s \gamma)_{\text{SM}} = (3.29 \pm 0.30) \times 10^{-4} \ ,
\label{bsgSM}
\end{equation}
where the theoretical error includes uncertainties from scale
dependence and standard model input parameters.

The most stringent experimental bounds are

\begin{eqnarray}
\text{CLEO: } && B(B\to X_s \gamma) = (3.15
\pm 0.35_{\text{stat}} \pm 0.32_{\text{syst}}
\pm 0.26_{\text{model}}) \times 10^{-4}~\cite{CLEO} \\
\text{ALEPH: } && B(B\to X_s \gamma) = (3.11
\pm 0.80_{\text{stat}} \pm 0.72_{\text{syst}})
\times 10^{-4}~\cite{ALEPH} \ ,
\end{eqnarray}
which may be combined in a weighted average of~\cite{KN}

\begin{equation}
B(B\to X_s \gamma)_{\text{exp}} = (3.14 \pm 0.48) \times 10^{-4} \ .
\label{bsgtotalexp}
\end{equation}
Bounds on SUSY parameter space are extremely sensitive to the
treatment of errors.  With this in mind, however, to guide the eye in
the figures below, we also include bounds from Eq.~(\ref{bsgtotalexp})
with $2\sigma$ experimental errors:

\begin{equation}
2.18 \times 10^{-4} < B(B\to X_s \gamma) < 4.10 \times 10^{-4} \ .
\label{bsgbound}
\end{equation}
Similar bounds would follow from combining 1$\sigma$ experimental and
theoretical errors linearly.

Given a set of parameters $\maux$, $m_0$, $\tb$, and $\sign (\mu)$, we
may now determine $B(B\to X_s \gamma)$, assuming the central value of
Eq.~(\ref{bsgSM}).  In Fig.~\ref{fig:bsghplus} we plot $B(B\to X_s
\gamma)$ as a function of $m_{H^{\pm}}$, for three representative
values of $\tb$, fixed choice of $\sign (\mu)$, and scanning over the
remaining parameters $\maux$ and $m_0$.  The solid lines show the
value when only the charged Higgs diagram is included.

\begin{figure}[tb]
\postscript{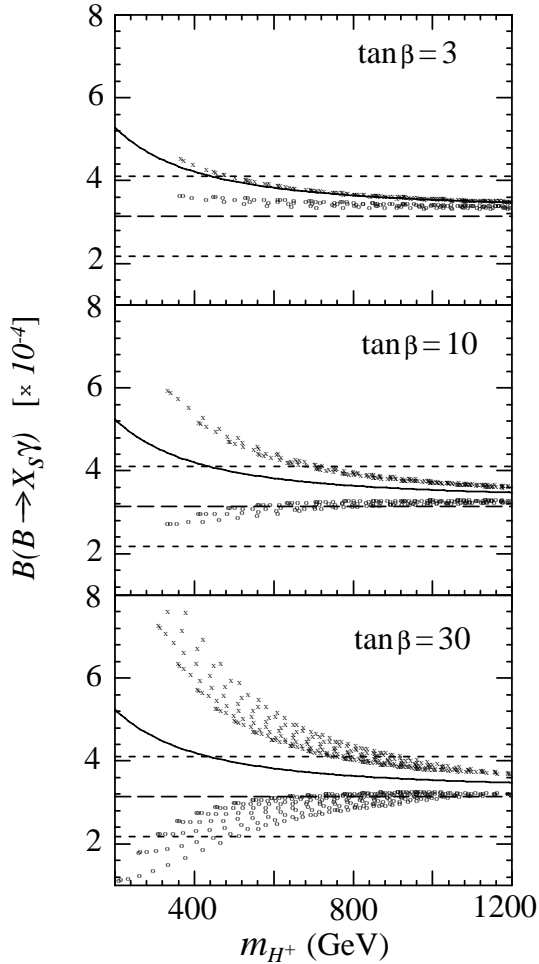}{0.44}
\caption{$B(B\to X_s \gamma)$, including standard model, $H^{\pm}$,
and SUSY contributions, as a function of $m_{H^{\pm}}$ for $\mu>0$
($\times$) and $\mu<0$ ($\circ$) and three representative values of
$\tb$.  Also shown are the theoretical prediction including only
the $\hplus$ contribution (solid), and the current experimental value
and $2\sigma$ experimental limits (dashed).}
\label{fig:bsghplus}
\end{figure}

As in minimal supergravity and gauge-mediated models, the neutralino
diagrams are negligible, but the chargino and, to a lesser extent,
gluino diagrams may be substantial, especially for large $\tb$.
In contrast to these other SUSY models, however, as a result of the
sign flips in $A_t$ and $M_3$ noted above, both chargino and gluino
contributions {\em enhance} the standard model prediction for $\mu>0$.
The parameter space with $\mu>0$ is thus highly constrained, and
requires large charged Higgs masses, especially for large $\tb$.
For example, for $\tb=30$, the upper bound of Eq.~(\ref{bsgbound})
implies $m_{\hplus}\agt 700\gev$, significantly more stringent than
the bound $m_{\hplus}\agt 400\gev$ that would apply in the absence of
chargino and gluino contributions.  For $\mu<0$, the supersymmetric
contributions may cancel the charged Higgs contribution, and the
parameter space is constrained only for very low $\maux$ and $m_0$,
where the destructive SUSY contributions push $B(B\to X_s \gamma)$
{\em below} experimental bounds.

In Figs.~\ref{fig:bsgp} and \ref{fig:bsgn} we plot $B(B\to X_s
\gamma)$ in the $(m_0, \maux)$ plane for various values of $\tb$ and
$\sign (\mu)$.  Regions excluded by Eq.~(\ref{bsgbound}) are shaded;
for $\mu > 0$ and large $\tb$, this includes much of the parameter
space with light sleptons and light Winos.

\begin{figure}[tb]
\postscript{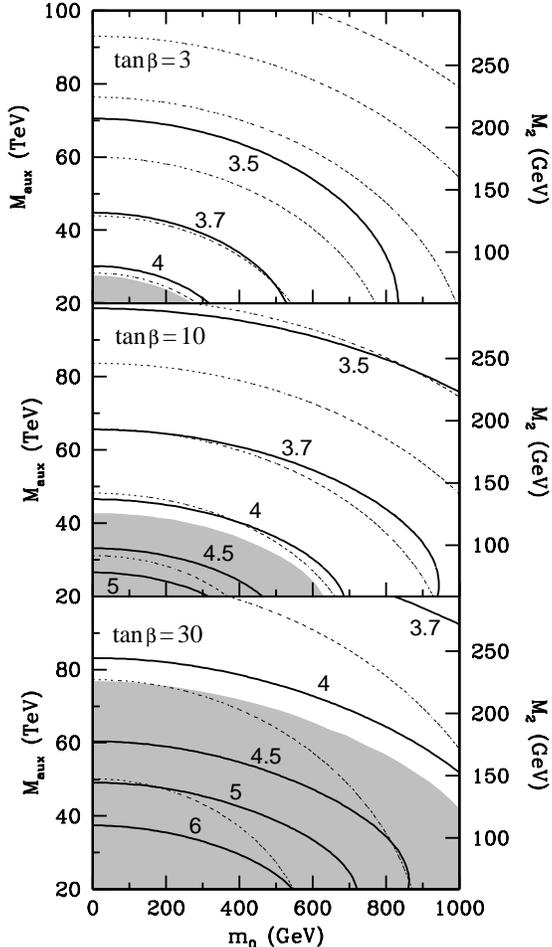}{0.44}
\caption{$B(B\to X_s \gamma)$ ($\times 10^{-4}$) (solid) for $\mu > 0$
and three representative values of $\tb$.  For reference, also shown
are contours of constant $m_{\hplus} = 500, 750, 1000, \cdots$ GeV,
from the lower left (dotted). The shaded region is excluded by $B(B\to
X_s \gamma) < 4.10 \times 10^{-4}$.}
\label{fig:bsgp}
\end{figure}

\begin{figure}[tb]
\postscript{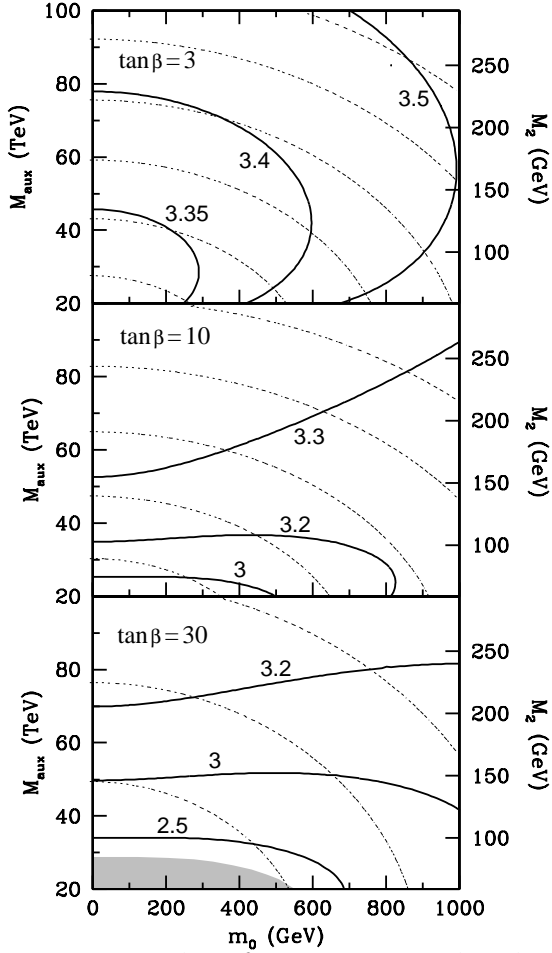}{0.44}
\caption{Contours as in Fig.~\ref{fig:bsgp}, but for $\mu<0$. The
shaded region is excluded by $B(B\to X_s \gamma) > 2.18 \times
10^{-4}$.}
\label{fig:bsgn}
\end{figure}

\subsection{Muon magnetic dipole moment}

While anomaly-mediated SUSY breaking does not contribute substantially
to flavor-violating observables involving the first and second
generations, it may give significant contributions to flavor-{\em
conserving} observables involving the first and second generations. It
is well-known that SUSY loops may give a sizable contribution to the
muon magnetic dipole moment (MDM)~\cite{g-2_susy}.  The
SUSY contribution to the muon MDM is from smuon-neutralino and
sneutrino-chargino loop diagrams.  Since these superparticles may have
masses comparable to the electroweak scale, these contributions may be
comparable to, or even larger than, electroweak contributions from
$W$- and $Z$-boson diagrams.  The on-going Brookhaven E821
experiment~\cite{e821} is expected to measure the muon MDM with an
accuracy of $0.4\times 10^{-9}$, which is about a few times smaller
than the electroweak contribution to the muon MDM.  Therefore, the
Brookhaven E821 experiment will provide an important constraint on
SUSY models.

In general, the muon anomalous MDM is given by the coefficient of the
``magnetic moment-type'' operator

\begin{eqnarray}
  {\cal L}_{\rm MDM} = \frac{e}{4m_\mu} a_\mu
  \bar{\mu} \sigma_{\mu\nu} \mu F_{\mu\nu} \ ,
\end{eqnarray}
where the anomalous magnetic moment $a_\mu$ is related to the muon
$g-2$ by $a_\mu = \frac{1}{2}(g-2)_\mu$.

As suggested from the structure of the operator, diagrams for the muon
anomalous MDM require a left-right muon transition. In SUSY diagrams,
this transition may occur through a chirality flip along the external
muon line, through left-right mixing in the smuon mass matrix, or
through the interaction of a muon and smuon with a Higgsino.  In the
latter two cases, the diagrams are proportional to the muon Yukawa
coupling constant and are therefore enhanced for large $\tb$.  These
diagrams also include gaugino mass insertions.  As a result, in the
large $\tb$ limit, the muon anomalous MDM is given by

\begin{eqnarray}
  a_\mu^{\rm SUSY} &\simeq&
  \frac{g_1^2}{16\pi^2} m_\mu^2 \mu M_1 \tb \times
  F_1(m_{\tilde{\mu}}^2,m_{\neutralino}^2)
 \nonumber \\ &&
+ \frac{g_2^2}{16\pi^2} m_\mu^2 \mu M_2 \tb \times
F_2(m_{\tilde{\mu}}^2, m_{\tilde{\nu}}^2, 
m_{\neutralino}^2, m_{\chargino}^2)\ ,
\end{eqnarray}
where the $F$ functions (see the last reference in
Ref.~\cite{g-2_susy}) are typically $F\sim m_{\rm SUSY}^{-4}$, with
$m_{\rm SUSY}$ being the mass scale of the superparticles in the loop.
For large $\tb$, then, the SUSY contribution $a_\mu^{\rm SUSY}$ is
approximately proportional to $\tb$ and may be much larger than the
electroweak contribution.

Results for the SUSY contribution to the muon MDM $a_\mu^{\rm SUSY}$
in the minimal anomaly-mediated model are given in Fig.~\ref{fig:amu}.
\begin{figure}[tb]
\postscript{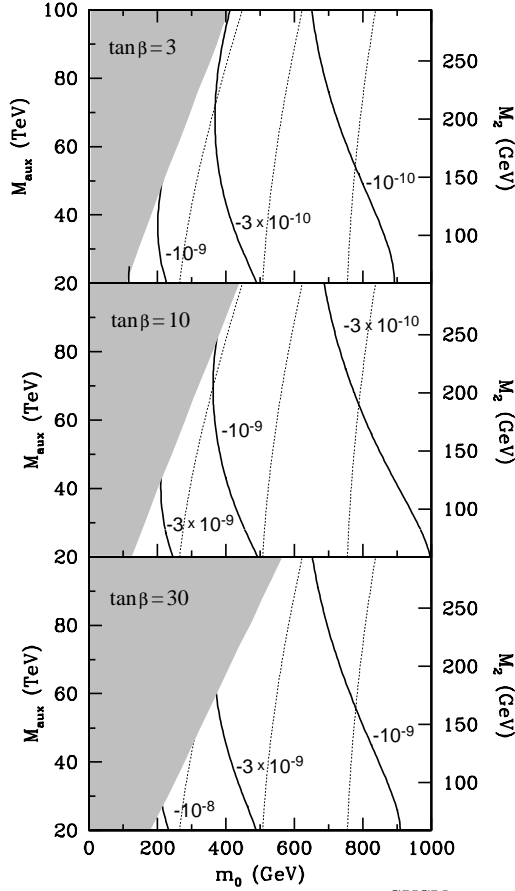}{0.41}
\caption{The muon anomalous magnetic moment $\amu = \frac{1}{2}
(g-2)^{\text{SUSY}}_{\mu}$ (solid) in the $(m_0, \maux)$ plane for
$\mu < 0$ and three representative values of $\tb$.  Also shown are
contours of constant $m_{\tilde{\mu}} = 250$, 500, and 750 GeV from
the left (dotted). The shaded region is excluded by $m_{\stau_1}>
70\gev$. Results for $\mu>0$ are of similar magnitude, but opposite in
sign.}
\label{fig:amu}
\end{figure}
Both $\tb$ enhanced and unenhanced contributions were included by
using the mass eigenstate bases of squarks, sleptons, neutralinos, and
charginos.  The SUSY contribution to the muon MDM is typically $\sim
10^{-8}-10^{-10}$, and is enhanced for large $\tb$.  Furthermore,
heavier superparticles suppress $a_\mu^{\rm SUSY}$, as expected.

Experimentally, the muon anomalous MDM is currently constrained to
be~\cite{PDG}
\begin{eqnarray}
  a_\mu^{\rm exp} = (1~165~923.0\pm 8.4)\times 10^{-9} \ ,
\end{eqnarray}
and hence the anomaly-mediated SUSY contribution is usually smaller
than the present experimental accuracy, unless $\tb$ is very large.
However, as mentioned above, in the near future, the Brookhaven E821
experiment will improve the measurement, with a projected error of
$0.4\times 10^{-9}$.  If this is realized, some anomaly may be seen in
the muon MDM in the anomaly-mediated SUSY breaking scenario,
particularly for moderate or large values of $\tb$.

\subsection{Electric dipole moments of the electron and neutron}
\label{sec:edm}

In general, parameters in SUSY models are complex, and (some
combinations of) their phases are physical.  In the anomaly-mediated
SUSY breaking scenario, most of the SUSY breaking parameters are
proportional to the single parameter $\maux$, and so many of the
phases can be rotated away.  In particular, the gaugino mass
parameters and the $A$ parameters can be made real simultaneously.
However, even in anomaly-mediated SUSY breaking, a physical phase may
exist in the $\mu$ and $B_\mu$ parameters since their origins are not
well-understood.  In our analysis, we have not assumed any relation
between $\mu$ and $B_\mu$, and have simply constrained them so that
electroweak symmetry is properly broken.  In this approach, one
physical phase remains, which is given by

\begin{equation} 
\theta_{\rm phys} \equiv {\rm Arg}(\mu B^*_\mu M_i) \ .
\end{equation}
If this phase is non-vanishing, electric dipole moments (EDMs) are
generated.  As is known from general analyses, the EDMs of the
electron and neutron may be extremely large unless $|\sin\theta_{\rm
phys}|$ is suppressed~\cite{edm_susy}.

To determine the constraints on this phase in the anomaly-mediated
framework, we calculate the electron and neutron EDMs with the minimal
anomaly-mediated model mass spectrum.  The EDM $d_f$ of a fermion $f$
is given by the effective electric dipole interaction

\begin{equation} 
{\cal L}_{\rm EDM} = -\frac{i}{2} d_f \bar{f}
\sigma_{\mu\nu} \gamma_5 f F_{\mu\nu} \ , 
\end{equation} 
which becomes ${\cal L}_{\rm EDM}\simeq d_f \vec{\sigma}\cdot \vec{E}$
in the non-relativistic limit.

The calculation of the electron EDM is similar to that of the muon
anomalous MDM, since the structure of the Feynman diagrams is almost
identical.  If the slepton masses are flavor universal, $a_\mu$ and
$d_e$ are approximately related by\footnote{In the calculation of the
muon anomalous MDM, we neglected the effect of CP violation.  If
$\sin\theta_{\rm phys}\neq 0$, $a_\mu$ is proportional to
$\cos\theta_{\rm phys}$ in the large $\tb$ limit.}

\begin{eqnarray}
  d_e \simeq \frac{m_e}{2m_\mu^2} \tan\theta_{\rm phys}
  \times a^{\rm SUSY}_\mu \ ,
\end{eqnarray}
Therefore, the electron EDM is also proportional to $\tb$.

The calculation of the up and down quark EDMs is also straightforward,
given the SUSY model parameters.  The only major difference from the
electron EDM is the contribution from the squark-gluino diagram.
However, in calculating the neutron EDM, we must adopt some model for
the structure of the neutron.  We use the simplest model, i.e., the
non-relativistic quark model.  The neutron EDM is then given by

\begin{eqnarray}
  d_n = \frac{1}{3} (4d_d - d_u) \ .
\end{eqnarray}
Since $d_d$ is also proportional to $\tb$, the neutron EDM is
also enhanced for large $\tb$.

Figures~\ref{fig:edme} and \ref{fig:edmn} show the EDMs of the
electron and neutron, respectively, in the minimal anomaly-mediated
model.  The EDMs are proportional to $\sin\theta_{\rm phys}$.  In
these plots, we assume maximal CP violation, i.e., $\sin\theta_{\rm
phys}=1$.

\begin{figure}[tb]
\postscript{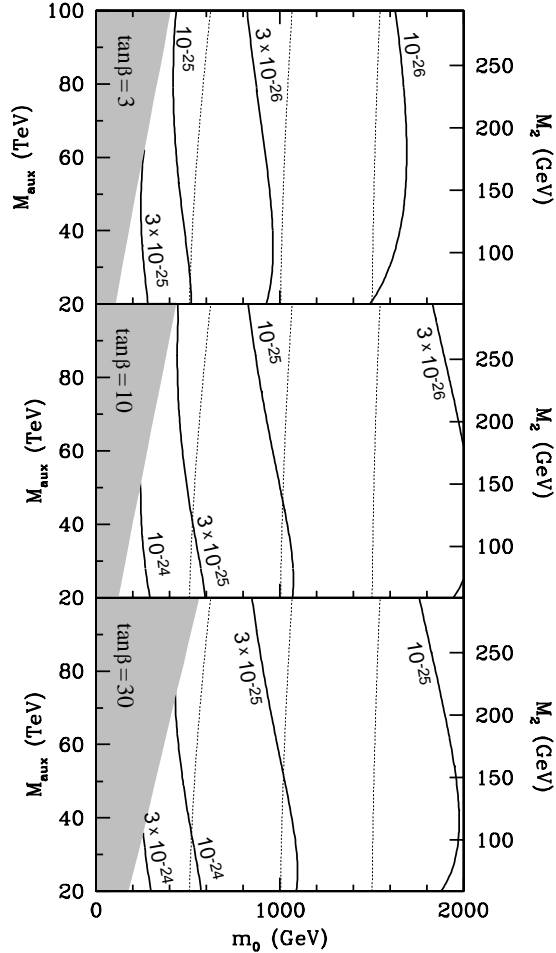}{0.44}
\caption{The electric dipole moment of the electron $|d_e|$ in units
of $e$-cm in the $(m_0, \maux)$ plane, assuming $\sin\theta_{\rm
phys}=1$, for $\mu < 0$ and three representative values of
$\tb$.  Also shown are contours of constant $m_{\selectron} =
250$, 500, and 750 GeV from the left (dotted). The shaded region is
excluded by $m_{\stau_1}> 70\gev$.}
\label{fig:edme}
\end{figure}

\begin{figure}[tb]
\postscript{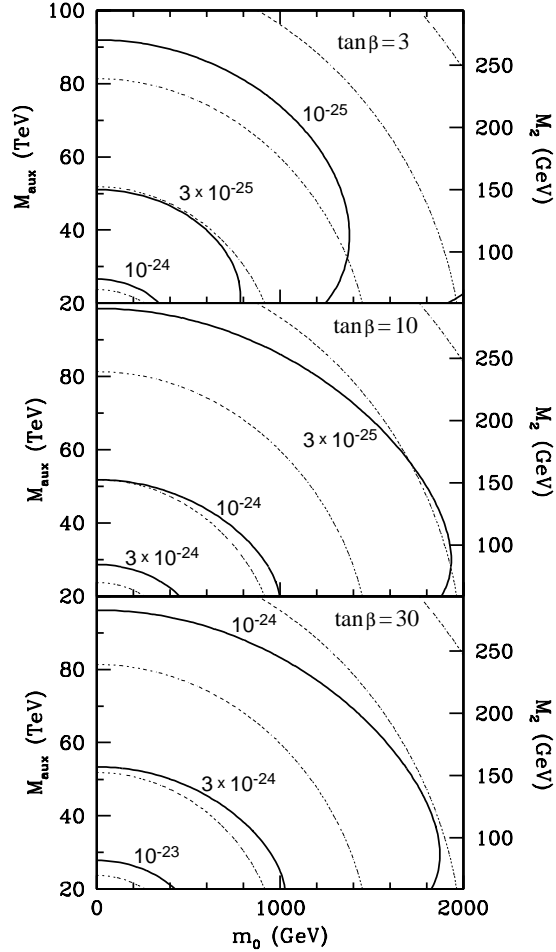}{0.44}
\caption{The electric dipole moment of the neutron $|d_n|$ in units
of $e$-cm in the $(m_0, \maux)$ plane, assuming $\sin\theta_{\rm
phys}=1$, for $\mu < 0$ and three representative values of
$\tb$.  Also shown are contours of constant $m_{\tilde{u}_L} =
500$, 1000,$\cdots$, 2500 GeV, from the lower left (dotted).}
\label{fig:edmn}
\end{figure}

Currently, there is no experimental result which suggests a
non-vanishing EDM, and experimental constraints on the EDMs are very
stringent.  For the electron EDM, using $d_e=(0.18 \pm 0.12 \pm 0.10)
\times 10^{-26}e$~cm~\cite{PRA50-2960}, we obtain the constraint
\begin{equation}
  |d_e| \leq 0.44 \times 10^{-26} e ~{\rm cm} \ ,
 \label{de_exp}
\end{equation}
where the right-hand side is the upper bound on $d_e$ at 90\%~C.L.
For the neutron, $d_n$ is constrained to be~\cite{PDG}
\begin{equation}
  |d_n| \leq 0.97 \times 10^{-25} e ~{\rm cm} \ .
 \label{dn_exp}
\end{equation}

The naturalness arguments of Sec.~\ref{sec:naturalness} play an
important part in evaluating the sensitivity of the EDMs.  For $d_e$
and small $\tb$, while very large effects are possible, $d_e$ may be
within the experimental bounds even for $|\sin\theta_{\rm phys}|$
close to 1 without violating the condition $|\mu| \alt 1 \tev$.  For
moderate and large $\tb$, $d_e$ becomes much larger, and the physical
phase $\theta_{\rm phys}$ is constrained to be $|\sin\theta_{\rm
phys}|\alt {\cal O} (10^{-2})$ for $m_0 \approx 1 \tev$.  However, for
such $\tb$, the naturalness bound on $m_0$ is also relaxed, and
reasonably large ${\cal O}(0.1)$ phases are possible in natural
regions of parameter space where $d_e$ is suppressed by slepton masses
of a few TeV.  Thus, while large effects comparable to current bounds
are predicted in much of parameter space, constraints from $d_e$ may
also be satisfied by superpartner decoupling in the minimal
anomaly-mediated model.  For $d_n$, similar conclusions hold.  In
fact, the constraints from $d_n$ on the CP-violating phases are more
easily satisfied, and $d_e$ appears to be the more stringent
constraint at present.

In our discussion, as noted above, we have not assumed a specific
model for the $\mu$ and $B_\mu$ parameters, and hence we regarded
$\theta_{\rm phys}$ as a free parameter.  However, several mechanisms
have been proposed to generate $\mu$ and $B_\mu$ in which
$\sin\theta_{\rm phys}$ vanishes~\cite{RS,PR,KSS}.  In those
scenarios, of course, $d_e$ and $d_n$ vanish, and the EDM constraints
are automatically satisfied.

\section{Conclusions}
\label{sec:conclusions}

In this study we have analyzed a model of ``supernatural
supersymmetry,'' in which squarks and sleptons may be much heavier
than their typical naturalness limits, and SUSY is broken in another
world.  SUSY breaking is then communicated to our world dominantly via
anomaly-mediation, and we have considered in detail a model in which
tachyonic sleptons are avoided by a non-anomaly-mediated universal
scalar mass $m_0$.

The novel naturalness properties of this model are a result of a
``focus point'' behavior in the RG evolution of $m_{H_u}^2$, such that
its weak scale value is highly insensitive to $m_0$.  Naturalness
bounds on superparticle masses are therefore highly variable and
differ from naive expectations.  Naturalness places strong bounds on
gaugino masses, and Wino masses $M_2 \alt 200\gev$ are preferred. On
the other hand, for moderate and large values of $\tb$, multi-TeV
values of $m_0$, and therefore slepton and squark masses, are natural.

A number of spectacular collider signals are possible.  The
possibility of a highly degenerate triplet of Wino LSPs has recently
attracted a great deal of attention~\cite{FMRSS,GGW,CDG,CDM,GM}. In
the minimal anomaly-mediated scenario, we find that Winos are not only
the LSPs in much of parameter space, but are typically light, with
mass $\alt 200 \gev$, and extraordinarily degenerate, with charged
Wino decay lengths of several centimeters.  Such Wino characteristics
are ideal for Tevatron searches, where Winos may appear as vertex
detector track stubs in monojet events.  The prospects for discovery
at the Tevatron in Run II or III are highly promising~\cite{FMRSS}.

In the remaining parameter space, the LSP is either the lighter stau,
or the tau sneutrino.  In the $\stau_1$ LSP scenario, the $\stau_1$ is
typically lighter than 200 GeV and is stable.  It may be found in
searches for stable charged massive particles at both
LEP~\cite{delphi} and the Tevatron~\cite{LEPwebpage,stauth,stauexp}.
In the $\tilde{\nu}_{\tau}$ LSP scenario, the Winos, $\stau_1$ and
sneutrinos are all $\alt 110 \gev$.  In both scenarios, ongoing
searches at LEP and the Tevatron will be able to probe substantial
portions of the relevant parameter space.

The minimal anomaly-mediated model also has a number of other features
that distinguish it from other models. In addition to characteristic
gaugino mass ratios, these include highly degenerate same-flavor
sleptons, and large left-right mixing.  If SUSY is discovered,
measurements of slepton masses and mixings will provide strong
evidence for or against the minimal model and its assumption of an
additional universal slepton mass.

We have also considered a variety of low energy observables that are
sensitive probes of anomaly-mediated parameter space.  Effects on the
flavor-changing process $\bsg$ may be large, and significant regions
of parameter space for large $\tb$ and $\mu>0$ are already excluded.
The anomalous magnetic moment of the muon may also be affected at
levels soon to be probed by experiment.  Finally, the electron and
neutron electric dipole moments provide rather strong constraints on
the CP-violation phase $\theta_{\text{phys}}$ in much of parameter
space, but even for large $\tb$, ${\cal O}(0.1)$ phases are still be
allowed for multi-TeV $m_0$ at its focus point naturalness limit.

It is interesting to note that positive signals in these low energy
experiments may not only provide evidence for SUSY, but may also
exclude some supersymmetric interpretations and favor others.  For
example, the signs of the SUSY contributions to $\bsg$ and $a_\mu^{\rm
SUSY}$ are determined by $\sign ( \mu M_3)$~\footnote{Here we assume
that the signs of $M_3$ and $A_t$ are correlated, as they are in
anomaly-mediation, and, through RG evolution, in gauge-mediated models
and minimal supergravity.} and $\sign ( \mu M_2)$, respectively.  A
large anomalous measurement of $a_\mu^{\rm SUSY}$ would imply large
$\tb$, and, given the current bounds on $\bsg$, a preferred sign for
$\sign (\mu M_3)$.  The sign of the $a_\mu^{\rm SUSY}$ anomaly then
determines $\sign (M_2 M_3)$.  For example, assuming a SUSY
interpretation, a large negative anomalous MDM measurement would imply
$M_2 M_3 < 0$, and would favor anomaly-mediated models over virtually
all other well-motivated models.

Finally, as stated in Sec.~\ref{sec:MAM}, the assumption of a
universal scalar mass contribution, while possibly generated by bulk
contributions~\cite{RS}, does not hold generally in anomaly-mediated
scenarios. Several features presented above depend on various parts of
this assumption, and we therefore close with a brief discussion of
these dependences.

The naturalness properties described above, and, in particular, the
focus point behavior, results from the fact that the
non-anomaly-mediated piece is identical for $m_{H_u}^2$, $m_{U_3}^2$,
and $m_{Q_3}^2$.  While the focus point mechanism as implemented here
relies on this subset of the universal boundary conditions, a variety
of other boundary conditions also have similar properties\footnote{For
example, the initial condition $(m_{H_u}^2, m_{U_3}^2, m_{Q_3}^2) =
m_0^2 (1,1+x,1-x)$, for any $x$, also leads to focus point behavior.},
and it would be interesting to explore applications of the focus point
mechanism in other settings.  The accidental degeneracy of left- and
right-handed sleptons, and the possibility for large left-right
mixings, holds only if both left- and right-handed sleptons receive
the same non-anomaly-mediated contribution. Measurement of large
left-right smuon mixing, along with confirmation of anomaly-mediated
gaugino mass parameters, for example, would therefore be strong
evidence for anomaly-mediation with a universal slepton mass
contribution. Finally, the low energy observables discussed are
sensitive quantitatively to either the hadronic or leptonic
superpartner spectrum.  However, qualitative results, such as the
stringency of constraints for large $\tb$, can be expected to remain
valid for a variety of anomaly-mediated models, as long as the
attractive flavor properties of anomaly-mediation are preserved in
these models and they do not have new large sources of flavor
violation.

\acknowledgements

We are grateful to Greg Anderson, Jon Bagger, Toru Goto, Erich
Poppitz, Lisa Randall, Yael Shadmi, Yuri Shirman, and Frank Wilczek
for helpful correspondence and conversations. The work of JLF was
supported by the Department of Energy under contract
DE--FG02--90ER40542 and through the generosity of Frank and Peggy
Taplin.  The work of TM was supported by the National Science
Foundation under grant PHY--9513835 and a Marvin~L.~Goldberger
Membership.

\appendix

\section*{Anomaly-Mediated Boundary Conditions}

In this appendix, we present the leading order soft supersymmetry
breaking terms, first for a general anomaly-mediated supersymmetric
theory, and then for the minimal anomaly-mediated model.

Consider a supersymmetric theory with simple gauge group $G$.  The
anomaly-mediated boundary conditions are completely specified in terms
of the gauge coupling $g$, supersymmetric Yukawa couplings

\begin{equation}
W = \frac{1}{6} Y^{ijk} \phi_i \phi_j \phi_k \ ,
\end{equation}
and the supersymmetry breaking parameter $\maux$.  

In the convention that the soft supersymmetry-breaking terms are
\begin{equation}
  {\cal L_{\text{SSB}}} = 
  -\frac{1}{2} M_\lambda (-i \lambda) (-i \lambda)
  - \frac{1}{2} (m^2)_i^j \tilde{\phi}^{\ast i}\tilde{\phi}_j
  -\frac{1}{6} A^{ijk} \tilde{\phi}_i \tilde{\phi}_j \tilde{\phi}_k \ ,
\label{convention}
\end{equation}
 the leading order anomaly-mediated soft supersymmetry breaking terms
are 
 \begin{eqnarray}
   M_\lambda|_{\rm AM} &=& \frac{1}{16\pi^2} b g^2 M_{\rm aux} \\
  {(m^2)}_i^j|_{\rm AM} &=& 
  \frac{1}{2} \left(\dot{\gamma}\right)_i^j M_{\rm aux}^2 \\
  A^{ijk}|_{\rm AM} &=& 
  - \sum_m \left( Y^{mjk} \gamma_m^i + Y^{imk} \gamma_m^j
  + Y^{ijm} \gamma_m^k\right) M_{\rm aux} \ ,
 \end{eqnarray}
where
\begin{equation}
\gamma_i^j= \frac{1}{2} Y_{imn} Y^{jmn} - 2\delta_i^j g^2 C(i) \ .
\end{equation}
Here $\dot{(\ )} \equiv d/d\ln\mu$, $Y_{ijk} = {Y^{ijk}}^\ast$, and
the one-loop $\beta$-function coefficient is $b = S(R) - 3 C(G)$,
where $C(i)$ is the quadratic Casimir invariant for representation
$i$, and $S(R)$ is the total Dynkin index summed over all the chiral
superfields. In terms of the matter field wavefunction $Z$,
$\gamma_i^j \equiv - \frac{1}{2} \dot{\left(\ln Z\right)}_i^j$.

For minimal field content, anomaly-mediated gaugino masses are given
as
\begin{equation}
  M_i = \frac{1}{16\pi^2} b_i g_i^2 M_{\rm aux} \ ,
\end{equation}
 where $b_i = (\frac{33}{5}, 1, -3)$ in the GUT normalization.
Furthermore, with the superpotential
 \begin{eqnarray}
  W = U_i \bold{Y_u}_{ij} Q_j H_u
  + D_i \bold{Y_d}_{ij} Q_j H_d
  + E_i \bold{Y_e}_{ij} L_j H_d \ ,
 \end{eqnarray}
 the flavor-dependent wavefunction factors are
 \begin{eqnarray}
  16\pi^2 \gamma_{H_u} &=& 3 \tr(\bold{Y_u}^\dagger \bold{Y_u})
  - \frac{3}{2}g_2^2 - \frac{3}{10}g_1^2 \\
  16\pi^2 \gamma_{H_d} &=& 3 \tr(\bold{Y_d}^\dagger \bold{Y_d})
  + \tr(\bold{Y_e}^\dagger \bold{Y_e})
  - \frac{3}{2}g_2^2 - \frac{3}{10}g_1^2 \\
  16\pi^2 \bold{\gamma_{Q}} &=& \bold{Y_u}^\dagger \bold{Y_u}
  + \bold{Y_d}^\dagger \bold{Y_d}
  - \frac{8}{3}g_3^2 - \frac{3}{2}g_2^2 - \frac{1}{30}g_1^2 \\
  16\pi^2 \bold{\gamma_{U}} &=& 2 \bold{Y_u}^* \bold{Y_u}^T
  - \frac{8}{3}g_3^2 - \frac{8}{15}g_1^2 \\
  16\pi^2 \bold{\gamma_{D}} &=& 2 \bold{Y_d}^* \bold{Y_d}^T
  - \frac{8}{3}g_3^2 - \frac{2}{15}g_1^2 \\
  16\pi^2 \bold{\gamma_{L}} &=& \bold{Y_e}^\dagger \bold{Y_e}
  - \frac{3}{2}g_2^2 - \frac{3}{10}g_1^2 \\
  16\pi^2 \bold{\gamma_{E}} &=& 2 \bold{Y_e}^* \bold{Y_e}^T
  - \frac{6}{5}g_1^2 \ ,
 \end{eqnarray}
 where the Yukawa couplings $\bold{Y}$ are $3\times 3$ matrices in
generation space.

The gauge and Yukawa coupling RG equations are as in Ref.~\cite{MV},
and are reproduced here for convenience and completeness:

\begin{eqnarray}
16\pi^2 \dot{g}_i &=& b_i g_i^3 \\
16\pi^2 \dot{\bold{Y_u}} &=& \bold{Y_u} \left[
3 \tr(\bold{Y_u} \bold{Y_u}^\dagger)+3\bold{Y_u}^\dagger\bold{Y_u}
+\bold{Y_d}^\dagger \bold{Y_d} -\frac{16}{3}g_3^2
- 3 g_2^2 - \frac{13}{15}g_1^2 \right] \\
16\pi^2 \dot{\bold{Y_d}} &=& \bold{Y_d} \left[
3 \tr(\bold{Y_d} \bold{Y_d}^\dagger)+
3 \tr(\bold{Y_e} \bold{Y_e}^\dagger)
+ 3\bold{Y_d}^\dagger\bold{Y_d}
+ \bold{Y_u}^\dagger \bold{Y_u} -\frac{16}{3}g_3^2
- 3 g_2^2 - \frac{7}{15}g_1^2 \right] \\
16\pi^2 \dot{\bold{Y_e}} &=& \bold{Y_e} \left[
3 \tr(\bold{Y_d} \bold{Y_d}^\dagger)
+ \tr(\bold{Y_e} \bold{Y_e}^\dagger)
+3 \bold{Y_e}^\dagger \bold{Y_e}
- 3 g_2^2 - \frac{9}{5}g_1^2 \right] \ .
\end{eqnarray}

Our sign convention for the $\mu$ and $A$ parameters is such that,
with soft terms as defined in Eq.~(\ref{convention}), the chargino
mass terms are $(\psi ^-)^T \bold{M_{\chargino}} \psi^+ + {\rm h.c.}$,
where $(\psi^{\pm})^T = (-i\tilde{W}^{\pm}, \tilde{H}^{\pm})$ and

\begin{equation}
\bold{M_{\chargino}} = \left( \begin{array}{cc}
 M_2                    &\sqrt{2} \, m_W \sin\beta  \\
\sqrt{2} \, m_W \cos\beta   &\mu            \end{array} \right) \ ,
\end{equation}
and the stop left-right mixing terms are $m_t (A_t-\mu \cot\beta)$.

\end{document}